\documentclass[11pt,preprint]{aastex}
\def\mbar{\ifmmode\overline{m}\else$\overline{m}$\fi}
\def\Mbar{\ifmmode\overline{M}\else$\overline{M}$\fi}
\def\mibar{\ifmmode\overline{m}_I\else$\overline{m}_I$\fi}
\def\MIbar{\ifmmode\overline{M}_I\else$\overline{M}_I$\fi}
\def\Nbar{\ifmmode\overline{N}\else$\overline{N}$\fi}

\def\ho{\ifmmode H_0\else$H_0$\fi}

\def\dmod{\ifmmode(m{-}M)_0\else$(m{-}M)_0$\fi}
\def\mM{\ifmmode(m{-}M)_0\else$(m{-}M)_0$\fi}
\def\vi{\ifmmode(V{-}I)\else$(V{-}I)$\fi}
\def\viz{\ifmmode(V{-}I)_0\else$(V{-}I)_0$\fi}
\def\EBV{\ifmmode E_{B-V}\else$E_{B-V}$\fi}

\shorttitle{Spectroscopy of High-$z$ Supernovae}
\shortauthors{Matheson et al.}

\begin{document}
\received{01 November 2004} 
\title{Spectroscopy of High-Redshift
Supernovae from the ESSENCE Project: The First Two
Years\altaffilmark{1}}

\author{Thomas Matheson,\altaffilmark{2}
St\'ephane Blondin,\altaffilmark{3}
Ryan J. Foley,\altaffilmark{4}
Ryan Chornock,\altaffilmark{4}
Alexei V. Filippenko,\altaffilmark{4}
Bruno Leibundgut,\altaffilmark{3}
R. Chris Smith, \altaffilmark{5}
Jesper Sollerman,\altaffilmark{6}
Jason Spyromilio,\altaffilmark{3}
Robert P. Kirshner,\altaffilmark{7}
Alejandro Clocchiatti,\altaffilmark{8} 
Claudio Aguilera,\altaffilmark{5}
Brian Barris,\altaffilmark{9}
Andrew C. Becker,\altaffilmark{10}
Peter Challis,\altaffilmark{7}
Ricardo Covarrubias,\altaffilmark{10}
Peter Garnavich,\altaffilmark{11}
Malcolm Hicken,\altaffilmark{7,12}
Saurabh Jha,\altaffilmark{4}
Kevin Krisciunas,\altaffilmark{11}
Weidong Li,\altaffilmark{4}
Anthony Miceli,\altaffilmark{10}
Gajus Miknaitis,\altaffilmark{13}
Jose Luis Prieto,\altaffilmark{14}
Armin Rest,\altaffilmark{5}
Adam G. Riess,\altaffilmark{15}
Maria Elena Salvo,\altaffilmark{16}
Brian P. Schmidt,\altaffilmark{16}
Christopher W. Stubbs,\altaffilmark{7,12}
Nicholas B. Suntzeff,\altaffilmark{5}
and John L. Tonry\altaffilmark{9}
}

\nopagebreak 

\altaffiltext{1}{\vspace{0.00cm}Based in part on observations obtained
  at the Cerro Tololo Inter-American Observatory, which is operated by
  the Association of Universities for Research in Astronomy,
  Inc. (AURA) under cooperative agreement with the National Science
  Foundation (NSF); the European Southern Observatory, Chile (ESO
  Programme 170.A-0519); the Gemini Observatory, which is operated by
  the Association of Universities for Research in Astronomy, Inc.,
  under a cooperative agreement with the NSF on behalf of the Gemini
  partnership: the NSF (United States), the Particle Physics and
  Astronomy Research Council (United Kingdom), the National Research
  Council (Canada), CONICYT (Chile), the Australian Research Council
  (Australia), CNPq (Brazil) and CONICET (Argentina) (Programs
  GN-2002B-Q-14, GN-2003B-Q-11, GS-2003B-Q-11); the Magellan
  Telescopes at Las Campanas Observatory; the MMT Observatory, a joint
  facility of the Smithsonian Institution and the University of
  Arizona; and the F.~L. Whipple Observatory, which is operated by the
  Smithsonian Astrophysical Observatory.  Some of the data presented
  herein were obtained at the W. M. Keck Observatory, which is
  operated as a scientific partnership among the California Institute
  of Technology, the University of California, and the National
  Aeronautics and Space Administration. The Observatory was made
  possible by the generous financial support of the W. M. Keck
  Foundation.}

\altaffiltext{2}{\vspace{0.00cm}National Optical Astronomy
  Observatory, 950 N.  Cherry Avenue, Tucson, AZ 85719-4933;
  {matheson@noao.edu}}

\altaffiltext{3}{\vspace{0.00cm}European Southern Observatory,
  Karl-Schwarzschild-Strasse 2, Garching, D-85748, Germany;
  {sblondin@eso.org}, {bleibund@eso.org}, {jspyromi@eso.org}}

\altaffiltext{4}{\vspace{0.00cm} Department of Astronomy,
  University of California, Berkeley, CA 94720-3411;
  {rfoley@astro.berkeley.edu}, {chornock@astro.berkeley.edu},
  {alex@astro.berkeley.edu}, {sjha@astro.berkeley.edu},
  {weidong@astro.berkeley.edu}}

\altaffiltext{5}{\vspace{0.00cm}Cerro Tololo Inter-American
  Observatory, Casilla 603, La Serena, Chile;
  {csmith@noao.edu}, {caguilera@ctio.noao.edu}, {arest@noao.edu}, 
  {nsuntzeff@noao.edu}}

\altaffiltext{6}{\vspace{0.00cm}Stockholm Observatory, AlbaNova,
  SE-106 91 Stockholm, Sweden; {jesper@astro.su.se}}

\altaffiltext{7}{\vspace{0.00cm}Harvard-Smithsonian Center for
  Astrophysics, 60 Garden Street, Cambridge, MA 02138;
  {kirshner@cfa.harvard.edu},{pchallis@cfa.harvard.edu},
  {mhicken@cfa.harvard.edu}, {cstubbs@fas.harvard.edu}}

\altaffiltext{8}{\vspace{0.00cm}Pontificia Universidad Cat\'{o}lica
  de Chile, Departamento de Astronom\'{i}a y Astrof\'{i}sica, Casilla
  306, Santiago 22, Chile; {aclocchi@astro.puc.cl}}

\altaffiltext{9}{\vspace{0.00cm}Institute for Astronomy, University
  of Hawaii, 2680 Woodlawn Drive, Honolulu, HI 96822;
  {barris@ifa.hawaii.edu}, {jt@ifa.hawaii.edu}}

\altaffiltext{10}{\vspace{0.00cm} Department of Astronomy, 
  University of Washington, Box 351580, Seattle, WA 98195-1580;
  {becker@darkstar.astro.washington.edu},
  {ricardo@astro.washington.edu}, {amiceli@astro.washington.edu}}

\altaffiltext{11}{\vspace{0.00cm} Department of Physics, 
    University of Notre Dame, 
    225 Nieuwland Science Hall, Notre Dame, IN 46556-5670;
  {pgarnavi@nd.edu}, {kkrisciu@nd.edu}}

\altaffiltext{12}{\vspace{0.00cm}Department of Physics, Harvard
  University, 17 Oxford Street, Cambridge MA 02138}

\altaffiltext{13}{\vspace{0.00cm} Department of Physics, 
  University of Washington, Box 351560, Seattle, WA 98195-1560;
  {gm@u.washington.edu}}

\altaffiltext{14}{\vspace{0.00cm} Department of Astronomy, 
  Ohio State University, 4055 McPherson Laboratory, 
  140 W. 18th Ave., Columbus,
  Ohio 43210; {prieto@astronomy.ohio-state.edu}}

\altaffiltext{15}{\vspace{0.00cm}Space Telescope Science Institute,
  3700 San Martin Drive, Baltimore, MD 21218; {ariess@stsci.edu}}

\altaffiltext{16}{\vspace{0.00cm}The Research School of Astronomy and
  Astrophysics, The Australian National University, Mount Stromlo and
  Siding Spring Observatories, via Cotter Rd, Weston Creek PO 2611,
  Australia; {salvo@mso.anu.edu.au}, {brian@mso.anu.edu.au}}

\begin{abstract}

We present the results of spectroscopic observations of targets
discovered during the first two years of the ESSENCE project.  The
goal of ESSENCE is to use a sample of $\sim$200 Type Ia supernovae
(SNe~Ia) at moderate redshifts $(0.2 \lesssim z \lesssim 0.8)$ to
place constraints on the equation of state of the Universe.
Spectroscopy not only provides the redshifts of the objects, but also
confirms that some of the discoveries are indeed SNe~Ia.  This
confirmation is critical to the project, as techniques developed to
determine luminosity distances to SNe~Ia depend upon the knowledge
that the objects at high redshift are the same as the ones at low
redshift.  We describe the methods of target selection and
prioritization, the telescopes and detectors, and the software used to
identify objects.  The redshifts deduced from spectral matching of
high-redshift SNe~Ia with low-redshift SNe~Ia are consistent with
those determined from host-galaxy spectra.  We show that the
high-redshift SNe~Ia match well with low-redshift templates.  We
include all spectra obtained by the ESSENCE project, including 52
SNe~Ia, 5 core-collapse SNe, 12 active galactic nuclei, 19 galaxies, 4
possibly variable stars, and 16 objects with uncertain identifications.

\end{abstract}


\section{Introduction}

The revolution wrought in modern cosmology using luminosity distances
of Type Ia supernovae (SNe~Ia) \citep{schmidt98, riess98,
perlmutter99, riess01, knop03, tonry03, barris2304, riess04b} relies
upon the fact that the objects so employed are, in fact, SNe of Type
Ia.  Although the light-curve shape alone is useful
\citep[e.g.,][]{barris04}, the only way to be sure of the true nature
of an object as a SN~Ia is through spectroscopy.  The calculation of
luminosity distances depends upon the high-redshift objects being
SNe~Ia so that low-redshift calibration methods can be employed.  The
classification scheme for SNe is based upon the optical spectrum near
maximum \citep[see][for a review of SN types]{filippenko97}, so
rest-wavelength optical spectroscopy is necessary to properly identify
SNe~Ia at high redshifts.  Despite this significance, relatively
little attention has been paid to the spectroscopy of the
high-redshift SNe~Ia, with some notable exceptions \citep{coil00}.
Other publications that include high-redshift SN~Ia spectra include
\citet{schmidt98}, \citet{riess98}, \citet{perlmutter98},
\citet{leibundgut01}, \citet{tonry03}, \citet{barris2304},
\citet{blondin04}, \citet{riess04b}, and \citet{lidman04}.

In addition to providing evidence for the acceleration of the
expansion of the Universe, it was recognized at an early stage that
high-redshift SNe~Ia could put constraints on the equation of
state for the Universe \citep{garnavich98}, parameterized as $w =
P/(\rho c^2)$, the ratio of the dark energy's pressure to its density.
To further explore this, the ESSENCE project was begun.  The ESSENCE
(Equation of State: SupErNovae trace Cosmic Expansion) project is a
five-year ground-based SN survey designed to place constraints on the
equation-of-state parameter for the Universe using $\sim$200 SNe~Ia
over a redshift range of $0.2 \lesssim z \lesssim 0.8$ \citep[see][for
a more extensive discussion of the goals and implementation of the
ESSENCE project]{miknaitis05, smith05}.

Spectroscopic identification of optical transients is a major
component of the ESSENCE project.  In addition to confirming some
targets as SNe~Ia, the spectroscopy provides redshifts, allowing the
derived luminosity distances to be compared with a given cosmological
model.  So many targets are discovered during the ESSENCE survey that
a large amount of telescope time on 6.5~m to 10~m telescopes is
required.  In the first two years of the program, we were fortunate
enough to have been awarded over 60 nights at large-aperture
telescopes.  Even with this much time, though, our resources were
insufficient to spectroscopically identify all of the potentially
useful candidates.  This remains the most significant limiting factor
in achieving the ESSENCE goal of finding, identifying, and following
the desired number of SNe~Ia with the appropriate redshift
distribution.

Nonetheless, spectroscopic observations of ESSENCE targets in the time
available have been successful, with almost fifty SNe~Ia clearly
identified, and several more characterized as likely SNe~Ia.  Other
identifications include core-collapse SNe, active galactic nuclei
(AGNs), and galaxies.  The galaxy spectra may include an unidentified
SN component.

This paper will describe the results of the spectroscopic component of
the first two years of the ESSENCE program.  Year One refers to our
2002 Sep-Dec campaign; Year Two was our 2003 Sep-Dec campaign.  In
Section \ref{target}, we describe the process of target selection and
prioritization.  Section \ref{obs} describes the technical aspects of
the observations.  We discuss target identification in Section
\ref{id}.  The summary of results in terms of types of objects and
success rates is given in Section \ref{results}.  In addition, we
present in Section \ref{results} all of the spectra obtained,
including those of the SNe~Ia (with low-redshift templates),
core-collapse SNe, AGNs, galaxies, stars, and objects that remain
unidentified.

\section{Target Selection\label{target}}

The ESSENCE survey uses the Blanco 4~m telescope at CTIO with the
MOSAIC wide-field CCD camera to detect many kinds of optical
transients \citep{smith05}. Temporal coverage helps to identify
solar-system objects such as Kuiper Belt Objects (KBOs) and asteroids.
Known AGNs and variable stars can also be eliminated from the possible
SN~Ia list.  The remaining transients are all potentially SNe.  They
are also faint, requiring large-aperture telescopes to obtain spectra
of the quality necessary to securely identify the object.  Exposure
times on 8-10~m telescopes are typically about half an hour, but can be as
much as two hours.  Such telescope time is difficult to obtain in
quantity, so not all of the detected transients can be examined
spectroscopically.  We apply several criteria to prioritize target
selection for spectroscopic observation.

The first step in sorting targets is based upon the spectroscopic
resources available.  The equatorial fields used for the ESSENCE
program are accessible from most major astronomical sites, so the main
concern with matching targets to spectroscopic telescopes is the
aperture size of the telescope.  The ESSENCE targets are generally in
the range $18 \lesssim m_R \lesssim 24$ mag.  When 8-10~m telescopes
are unavailable, the fainter targets become lower in priority.  The
limit for low-dispersion spectroscopy to identify SNe with the 6.5~m
telescopes is $m_R \approx 22-23$ mag, although this will vary with
weather conditions and seeing.  If the full range of telescopes is
available, then targets are prioritized by magnitude for observation
at a given telescope.  The longitudinal distribution of spectroscopic
resources can be important if confirmation of a high-priority target
is made during a night when multiple spectroscopic resources are
available.  By the time a target is confirmed, the fields may have set
for telescopes in Chile, while they are still accessible from Hawaii.
This requires active, real-time collaboration between the group
finding SN candidates and those running the spectroscopic
observations.

One advantage of the ESSENCE program is that fields are imaged in
multiple filters, allowing for discrimination of targets by color.
\citet{tonry03} present a table of expected SN~Ia peak magnitudes as a
function of redshift; see also \citet{poznanski02}, \citet{galyam04},
\citet{riess04a}, \citet{strolger04}, and \citet{smith05} for
discussions of color selection for SN candidates.  Given apparent
$R$-band and $I$-band magnitudes, one can calculate the $R-I$ color
and compare that with an expected color for those magnitudes.  The
cadence of the ESSENCE program (returning to the same field every four
days) will likely catch SNe at early phases (i.e., before maximum
brightness).  Early core-collapse SNe are bluer than SNe Ia, as are
AGNs.  For example, when selecting for higher-redshift targets,
objects with $R-I \lesssim 0.2$ mag were considered unlikely to be
SNe~Ia, while objects with $R-I \gtrsim 0.4$ mag were made high
priority for spectroscopic observation.  The exact values of $R-I$
used for selection depended on the observed $R$-band magnitude.  This
method was used more consistently in the last month of Year Two,
reducing the fraction of spectroscopic targets that were identified as
AGNs from $\sim10\%$ over the lifetime of the project to $\sim5\%$
during that month.

The cadence of the ESSENCE program is designed to catch SNe early.  At
the start of an observing campaign or after periods of bad weather,
though, we may have missed SNe during their rise to maximum brightness
and only caught them while they are declining from maximum.  If a
target is brightening, then it is a higher priority than one that is
not.  This prioritization by phase of the SN became even more
important when our {\it Hubble Space Telescope} ($HST$) program to
observe some of the ESSENCE SNe~Ia was active
\citep[see][]{krisciunas05}.  The response time of $HST$ for a new
target, even if the rough position on the sky is known from our chosen
search fields, is still on the order of several days.  To ensure that
$HST$ was not generally looking at SNe~Ia after maximum brightness, we
would emphasize targets for spectroscopic identification that appeared
to be at an early epoch.  In addition, we chose fainter objects, as
higher-redshift SNe~Ia were a prime motivation for $HST$ photometry.
The $HST$ observations, while still targets of opportunity, were
scheduled for specific ESSENCE search fields, so new targets in those
fields were given the highest priority.

The position of the SN in the host galaxy also influences the priority
for observation.  An optical transient located at the core of a galaxy
is often an AGN, rather than an SN.  The color selection described
above is a less-biased predictor.  In addition, even if the object is
an SN, the signal of the SN itself is diluted by the light of the
galaxy, making proper identification difficult.  Objects that are
well-separated from the host galaxy are given a higher priority.
Being too far from the galaxy can, however, present another
problem---the difficulty in obtaining a spectrum of the host in
addition to the SN.  Without a high signal-to-noise ratio (S/N)
spectrum of the host, there is no precise measure of the redshift.
This is especially true if the host galaxy cannot be included in the
slit with the target, either to orient the slit at the parallactic
angle or as a result of other observational constraints.  In addition,
host galaxies can be faint, so the large luminosity contrast with the
SN makes detection of the host problematic (the so-called ``hostless''
SNe), although we did not reject any candidates solely for this
reason.  The best compromise is to have an object well separated from
the host, but with the host still in the spectrograph slit.  Without
narrow-line features from the host (either emission or absorption
lines), the redshift can be difficult to determine.  This lack of a
host-galaxy spectrum became less of a concern, though, as we found
that the SN spectrum itself is a relatively accurate, if less precise,
measure of the redshift (see discussion below).  The light curve alone
can be used to estimate distances in a redshift-independent way
\citep{barris04}, but only with a well-sampled and accurate light
curve.

The target selection process is complex and dynamic.  Biases are
introduced by some of the steps; for example, SN candidates near the 
centers of galaxies are less likely to be observed.  Since the goal is to
optimize the spectroscopic telescope time to identify SNe~Ia in a
specific redshift range, we have chosen these selection processes as
our best compromise.  The biases introduced may make the sample of
SNe~Ia identified problematic for uses in statistical studies of the
nature of SNe~Ia at high redshift.

\section{Observations\label{obs}}

Spectroscopic observations of ESSENCE targets were obtained at a wide
variety of telescopes: the Keck I and II 10~m telescopes, the VLT 8~m
telescopes, the Gemini North and South 8~m telescopes, the Magellan
Baade and Clay 6.5~m telescopes, the MMT 6.5~m telescope, and the
Tillinghast 1.5~m telescope at the F.~L.~Whipple Observatory (FLWO).
The spectrographs used were LRIS \citep{oke95} with Keck I, ESI
\citep{sheinis02} with Keck II, FORS1 with VLT \citep{appenzeller98},
GMOS \citep{hook02} at Gemini (North and South), IMACS
\citep{dressler04} with Baade, LDSS2 \citep{mulchaey01} with Clay, the
Blue Channel \citep{schmidt89} at MMT, and FAST \citep{fabricant98} at
FLWO.  Nod-and-shuffle techniques \citep{glaze01} were used with GMOS
(North and South) and IMACS to improve sky subtraction in the red
portion of the spectrum.

\addtocounter{footnote}{16}

Standard CCD processing and spectrum extraction were accomplished with
IRAF\footnote{IRAF is distributed by the National Optical Astronomy
Observatory, which is operated by the Association of Universities for
Research in Astronomy, Inc., under cooperative agreement with the
National Science Foundation.}.  Most of the data were extracted using
the optimal algorithm of \citet{horne86}; for the VLT data, an
alternative extraction method based upon Richardson-Lucy restoration
\citep{blondin04} was employed.  Low-order polynomial fits to
calibration-lamp spectra were used to establish the wavelength scale.
Small adjustments derived from night-sky lines in the object frames
were applied.  We employed IRAF and our own IDL routines to flux
calibrate the data and, in most cases, to remove telluric lines using
the well-exposed continua of the spectrophotometric standards
\citep{wade88,matheson00}.

\section{Target Identification\label{id}}

Once a calibrated spectrum is available, the next step is to properly
classify the object.  For brighter objects that yield high S/N
spectra, an SN is often easy to distinguish and classify.  Most of the
ESSENCE targets are faint enough to be difficult objects even for
large-aperture telescopes.  The resulting noisy spectra can be
confusing.  Even for well-exposed spectra, though, exact
classification can occasionally still be challenging.

For SNe, the classification scheme is based upon the optical spectrum
near maximum brightness \citep{filippenko97}.  Type II~SNe are distinguished by
the presence of hydrogen lines.  The Type I~SNe lack hydrogen, and are further
subdivided by the presence or absence of other features.  The hallmark
of SNe~Ia is a strong \ion{Si}{2} $\lambda6355$ absorption
feature.  Near maximum brightness, this absorption is blueshifted by
$\sim$10,000 km~s$^{-1}$ and appears near 6150~\AA.  In SNe~Ib,
this line is not as strong, and the optical helium series dominates
the spectrum.  The SNe~Ic lack all these identifying lines.

At high redshift, the \ion{Si}{2} $\lambda6355$ feature is at
wavelengths inaccessible to optical spectrographs, so the
identification relies upon the pattern of features in the rest-frame
ultraviolet (UV) and blue-optical wavelengths.  The \ion{Ca}{2} H\&K
$\lambda\lambda3934, 3968$ doublet is a distinctive feature in SNe~Ia,
but it is also present in SNe~Ib/c, so the overall pattern is
important for a clear identification as a SN~Ia.  Other important
features to identify SNe~Ia include \ion{Si}{2} $\lambda4130$,
\ion{Mg}{2} $\lambda4481$, \ion{Fe}{2} $\lambda4555$, \ion{Si}{3}
$\lambda4560$, \ion{S}{2} $\lambda4816$, and \ion{Si}{2} $\lambda5051$
\citep[see, e.g.,][]{coil00, jeffery92, kirshner93, mazzali93}.

The first stage of classification is done by eye.  Drawing upon the
extensive experience of the spectroscopic observers associated with
ESSENCE, we can provide a solid evaluation of the spectrum.  Objects
such as AGNs and normal galaxies are fairly easy to distinguish.  The
SNe~Ia are also often clear, but some fraction of the data will
require more extensive analysis.  The first step is simply to make
certain that the collective expertise is used, rather than just the
individual at the telescope.  Spectroscopic results are widely
disseminated via e-mail and through an internal web page, allowing
rapid examination of any questionable spectrum by the entire
collaboration.  Broad discussion often leads to a consensus.

In addition to the traditional by-eye approach, we employ automated
comparisons.  If the object is likely to be a SN~Ia, and if the S/N
ratio is sufficiently high and the rest-wavelength coverage appropriate, 
we can use a spectral-feature aging routine \citep{riess97} that 
compares specific components of the SN~Ia spectrum with a library of
SN~Ia spectra at known phases.  This can pin down the epoch of a
SN~Ia to within a few days.  This program, though, is limited to
normal SNe~Ia (i.e., not spectroscopically peculiar objects, which are 
often overluminous or underluminous).  In addition, it does not identify 
objects that do not match the Type Ia~SN spectra in the library.

For a more general identification routine, we use an algorithm called
SuperNova IDentification \citep[SNID; ][]{tonry05}.  This program
takes the input spectrum and compares it against a library of objects
of many types.  The templates include SNe~Ia of various luminosity
classes and at a range of ages, core-collapse SNe, and galaxies.  The
offset in wavelength caused by redshift is a free parameter, so the
output includes an estimate of the redshift of the object.  The
routine compares the input with the library and returns the most
likely match.  The comparison is weighted by the amount of overlap
between the input spectrum and the template.  For a subset of the
objects ($\lesssim 10\%$), the SNID comparison is not optimal.  This
may be the result of contamination by galaxy light, the lack of a
matching template in the SNID library, poor S/N of the spectrum in
question, or a problem in the routine.  All SNID comparisons are
checked by eye for a qualitative judgment of the goodness of fit.

The redshift of the object can also be directly determined from the
spectrum itself if narrow emission or absorption lines associated
with the host galaxy are present.  Occasionally, observations are set
up to include the host galaxy in the spectrograph slit specifically
for the purpose of obtaining a redshift.  If there is a strong enough
signal of a galaxy spectrum, but no clearly identifiable narrow
emission or absorption lines, cross-correlation with an absorption
template could be used.  For the spectra that had narrow emission or
absorption lines (or were cross-correlated with a template), we report
the redshift to three significant digits.  If the redshift
determination is based solely on a comparison of the SN spectrum to a
low-redshift analog, the redshift is less certain, and we only report
the value to two significant digits.

\begin{figure}
\plotone{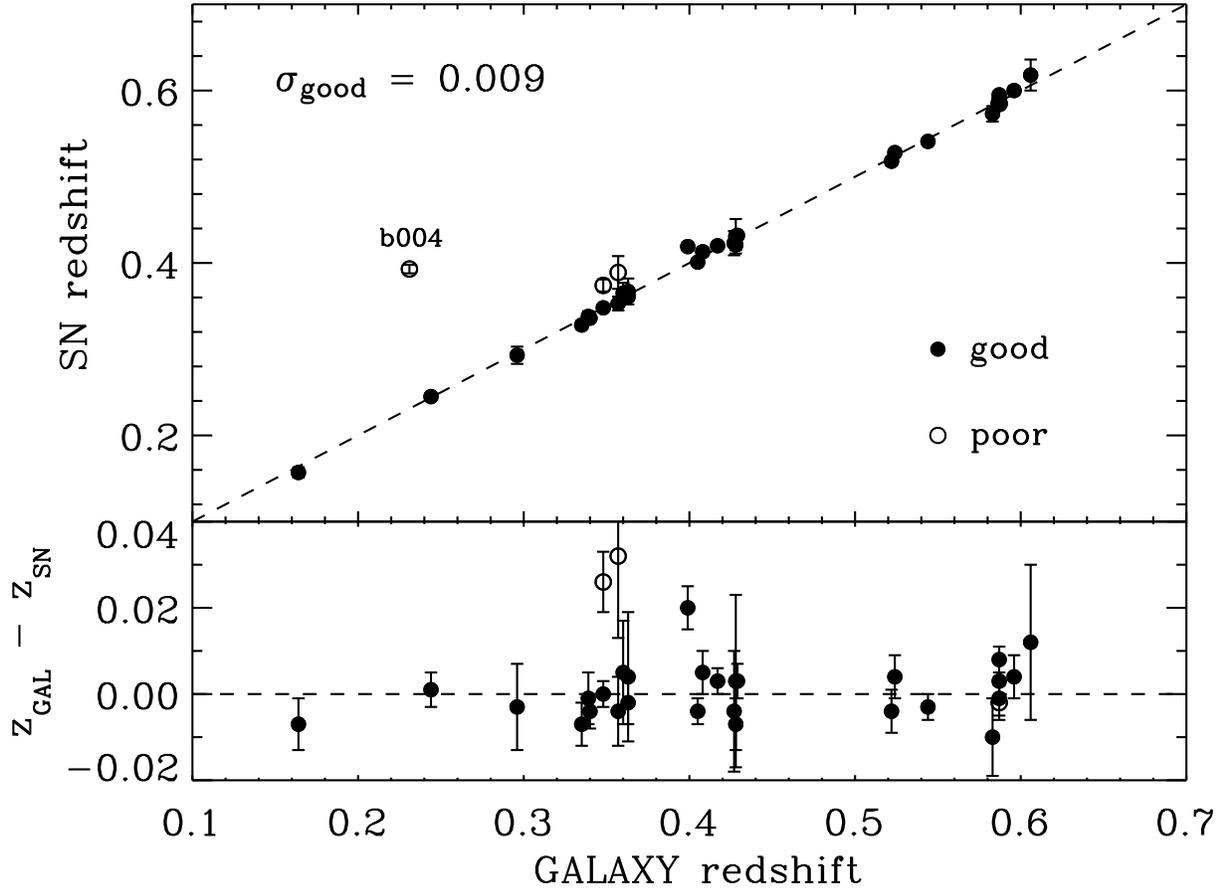}
\caption{Comparison of redshifts as determined by SNID and from narrow
emission or absorption lines in the host-galaxy spectrum.  Qualitative
grades for the fits in SNID are assigned and the good fits
(\emph{solid circles}) are shown as well as the poor fits (\emph{open
circles}). The dispersion around one-to-one correspondence of the
redshifts for the good data is excellent, with $\sigma=0.009$ (when no
errors are assumed for the SNID output).  There is one outlier (b004)
for which the redshift determination using SNID is highly degenerate
as it is likely to be a peculiar SN~Ia (see text); we do not show b004
in the residual plot for the sake of clarity.  The error bars in the
lower plot correspond to $\sigma_{\rm good}$.  Note that the mean
residual is $\sim 10^{-4} \ll \sigma_{\rm good}$, which shows that
there are no systematic effects associated with the use of SNID in
determining the SN redshift.\label{snidfig}}
\end{figure}

For the objects with a more precise redshift derived from the host
galaxy, we can compare the galactic redshift with the value of the
redshift estimated by SNID.  Figure \ref{snidfig} shows that the SNID
redshifts agree well with the galaxy redshifts.  Thus, for objects
without precise redshifts from host galaxy spectra, the SNID redshifts
can be used as reliable substitutes.  In the cases where SNID did not
agree with a galaxy redshift, we forced the redshift to match in order
to find the best-fitting template, but all the supernova-based
redshifts reported in this paper correspond to the ``un-forced'' SNID
result.

\section{Results\label{results}}

The results of our spectroscopic observations during the first two
years of the ESSENCE program are summarized in Table
\ref{resulttable}.  There are 46 SNe~Ia (and 5 additional likely
SNe~Ia), along with 5 core-collapse SNe.  Note also that there were 54
transients in the first two years that were not observed
spectroscopically.  Through the target selection methods described in
Section \ref{target}, we were able to prioritize the more likely
candidates, but many of these were not observed solely because of the
lack of sufficient spectroscopic resources.  This became more of an
issue toward the end of Year Two, when good weather and increasingly
efficient detection algorithms increased the number of transients
discovered.

\begin{figure}
\plotone{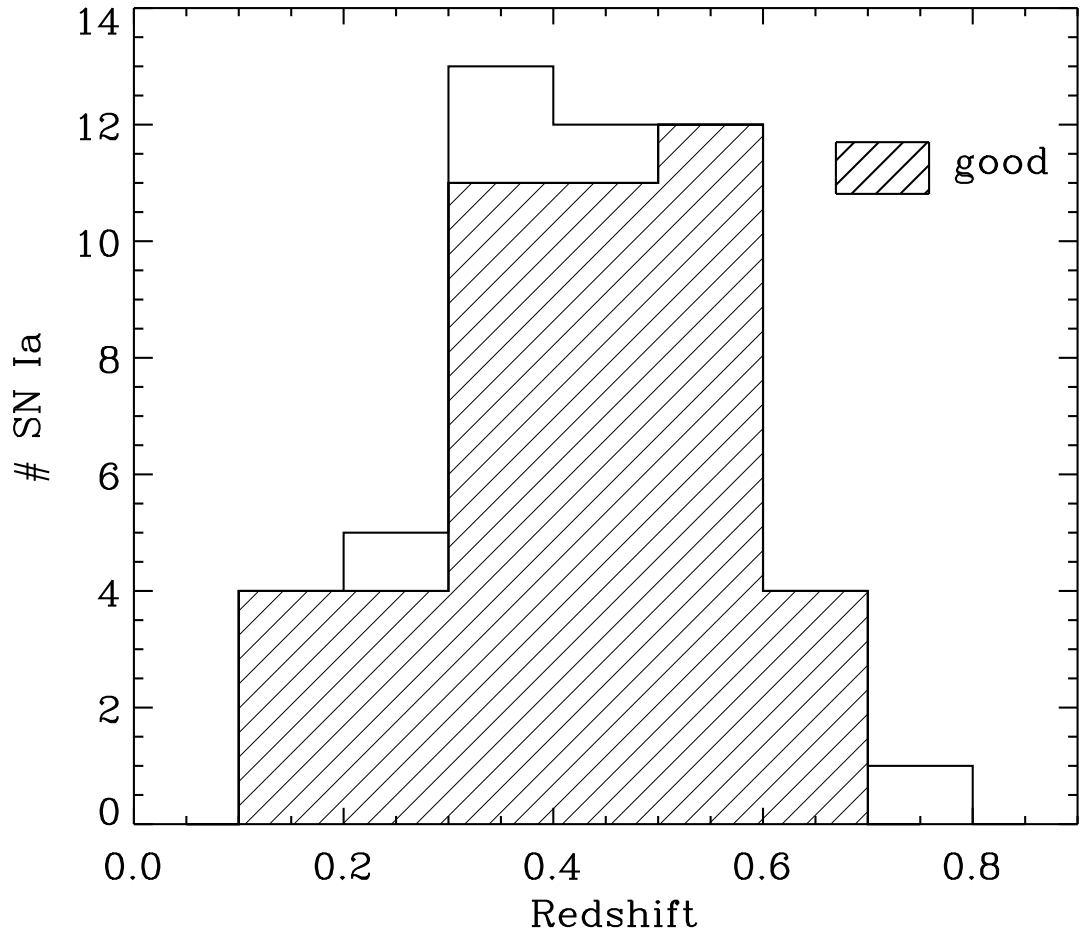}
\caption{Redshift distribution of spectroscopically identified SNe~Ia
  from the first two years of the ESSENCE project.  The SNe for which
  we judge that the SNID fit is good are indicated by the
  cross-hatched area, while those that were poor fits are indicated by
  the blank spaces.\label{zdistfig}}
\end{figure}

The goal of the ESSENCE project is to find $\sim$200 SNe~Ia over
the redshift range $0.2 \lesssim z \lesssim 0.8$.  In Figure
\ref{zdistfig}, we show the actual distribution in redshift of the
SNe~Ia from the ESSENCE project that are spectroscopically
confirmed.  There are SNe over the entire targeted redshift range,
although there are fewer at the high end ($z \gtrsim 0.6$).  A
significant fraction of the signal of $w$ is accessible at $z \approx
0.5$ \citep{miknaitis05}, but a goal for the last three years of the
program is to ensure that the SNe~Ia observed spectroscopically
are distributed optimally over our targeted redshift range.  This
highlights the importance of the 8-10~m telescopes such as
Gemini, the VLT, and Keck that are critical to spectroscopy of the
faint objects at the high-redshift end of our range.

\begin{figure}
\plotone{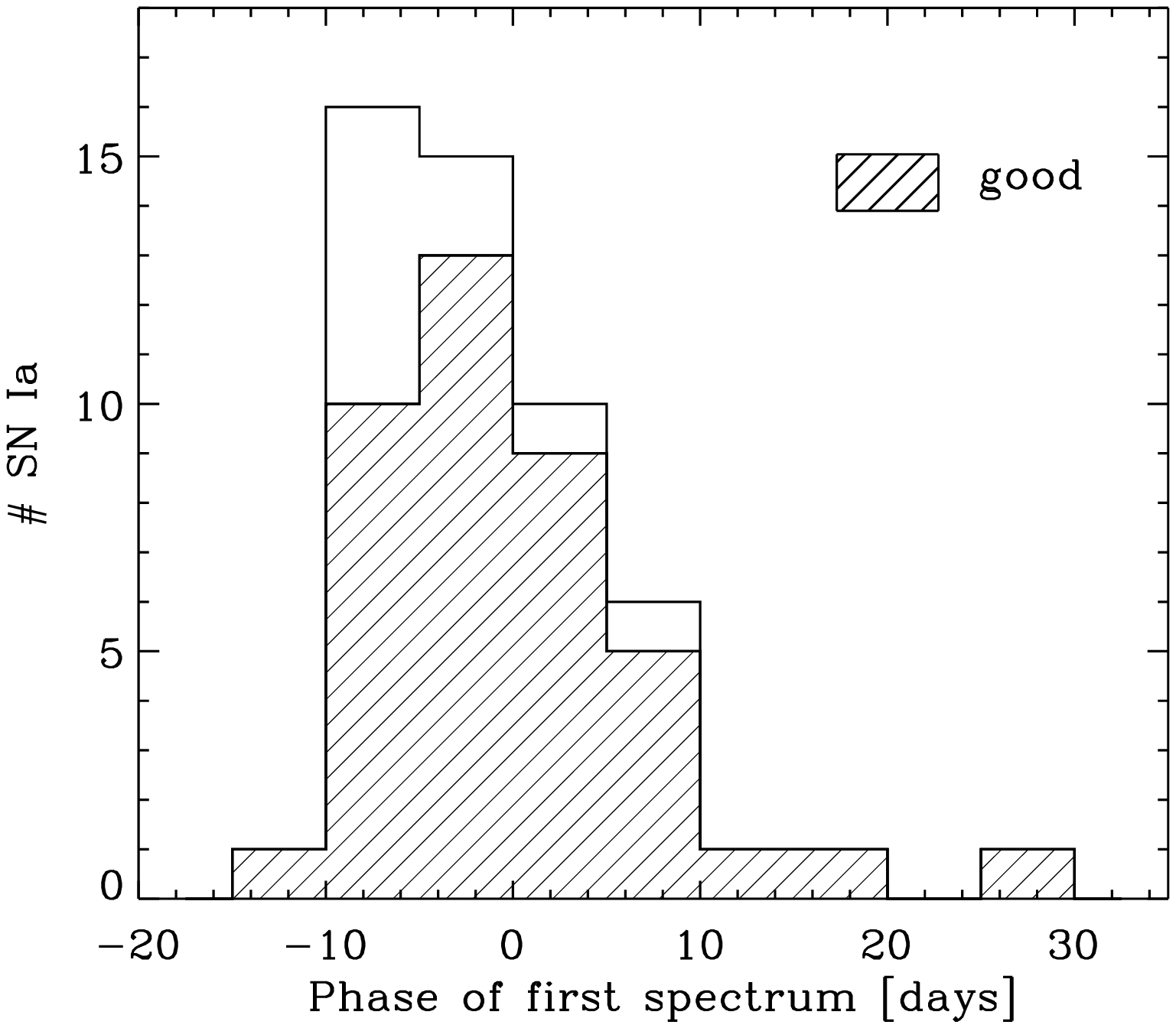}
\caption{Age distribution (relative to maximum brightness) of
  spectroscopically identified SNe~Ia from the first two years of
  the ESSENCE project.  Ages are determined from spectroscopic
  features alone.  The SNe for which we judge that the SNID fit is
  good are indicated by the cross-hatched area, while those that were
  poor fits are indicated by the blank spaces.  For objects with
  multiple epochs of spectroscopy, this figure only reflects the first
  spectrum.\label{agedistrib}}
\end{figure}

Both SNID and the spectral-feature aging method described in Section
\ref{id} give an indication of the age of the SN~Ia.  Light curves
will provide a more precise measure of the age of the SN at the time
of the spectroscopy, but an estimate of the epoch of the spectrum to
within a few days is possible from the spectral features alone.
Figure \ref{agedistrib} shows the distribution in age (relative to
maximum brightness) at the time of spectroscopy (not discovery, as
spectra are often taken up to several days after discovery).  In the
15 cases\footnote{These are b008, b010, b013, b020, b022, b023, b027,
c003, c012, c015, d086, d093, e029, e108, and f076.} where we have
spectra of the same SN~Ia at multiple epochs, the relative ages are
consistent with the times of the spectroscopic exposures (also
considering the effects of cosmological time dilation and probable
errors of the fits of $\sim\pm3$ days).  There is one exception to
this consistency (b027), but at later epochs when the spectra are
changing less.

Table \ref{sntable} is a list of all ESSENCE targets that were
selected for spectroscopic identification.  The results for these
first two years include 52 SNe~Ia or likely SNe~Ia (Figure 4), 4
SNe~II (Figure \ref{IIfig}), 1 SN~Ib/c (Figure \ref{IIfig}), 12 AGNs
(Figure 6), 4 possibly variable stellar objects (Figure
\ref{starfig}), 19 galaxies (Figure 8), and 16 objects of unknown
classification (Figure 9).  There were 10 objects for which we pointed
the telescope at the target and did not get a spectrum, either because
of poor sky conditions or the target was actually a solar-system
object and had moved out of the field.

\begin{figure}
\figurenum{4a}
\epsscale{0.9}
\plotone{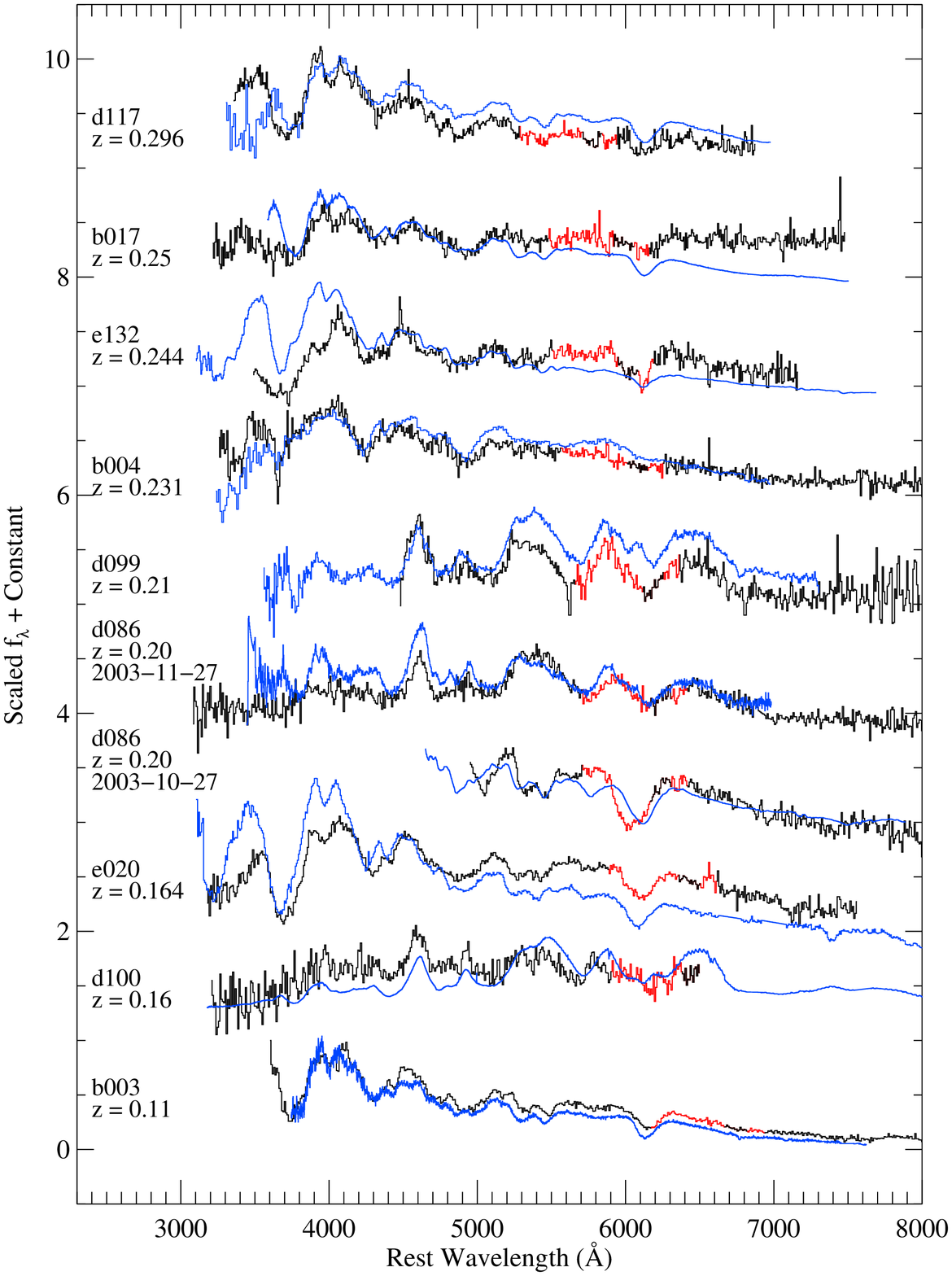}
\caption{Rest-wavelength spectra of SNe~Ia (or likely SNe~Ia) from the
  first two years of the ESSENCE project in order of increasing
  redshift.  Each ESSENCE SN (\emph{black line}) is overplotted by a
  low-redshift SN~Ia (\emph{blue line}) for comparison.  In addition,
  each spectrum is labeled with the ESSENCE identification number and
  the deduced redshift.  Spectra of the uncertain SNe~Ia are indicated
  with an asterisk (*).  The deredshifted regions of the spectra that
  are strongly affected by atmospheric absorption are shown in red.
  The flux scale is $f_{\lambda}$ with arbitrary additive offsets
  between the spectra.\label{Iafig}}
\end{figure}
\begin{figure}
\figurenum{4b}
\plotone{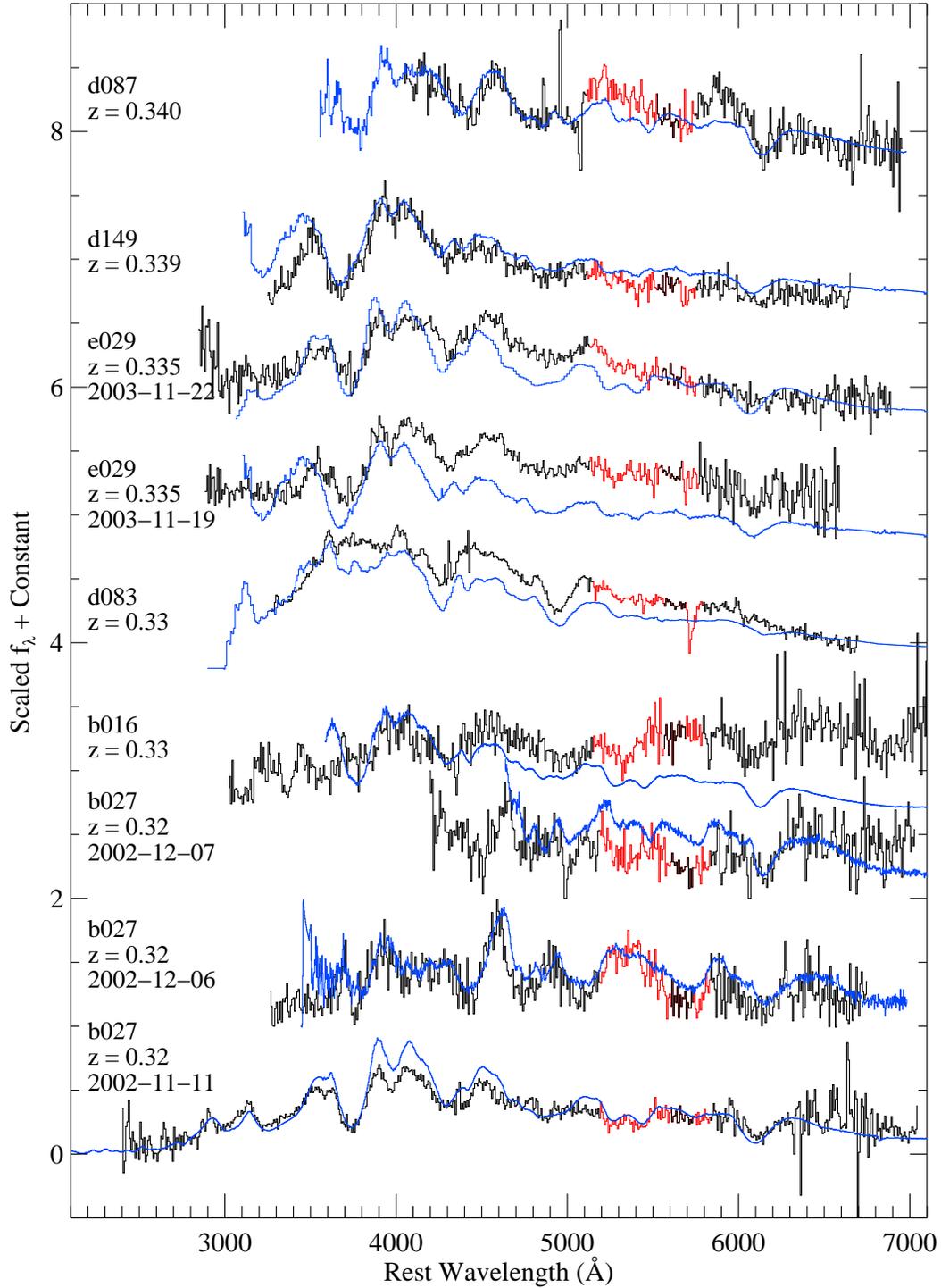}
\caption{Rest-wavelength spectra of ESSENCE SNe~Ia as in Figure 4a.
  The 2003-01-03 spectrum of c012 is a weighted average of the Clay
  and GMOS spectra.
  }
\end{figure}
\begin{figure}
\figurenum{4c}
\plotone{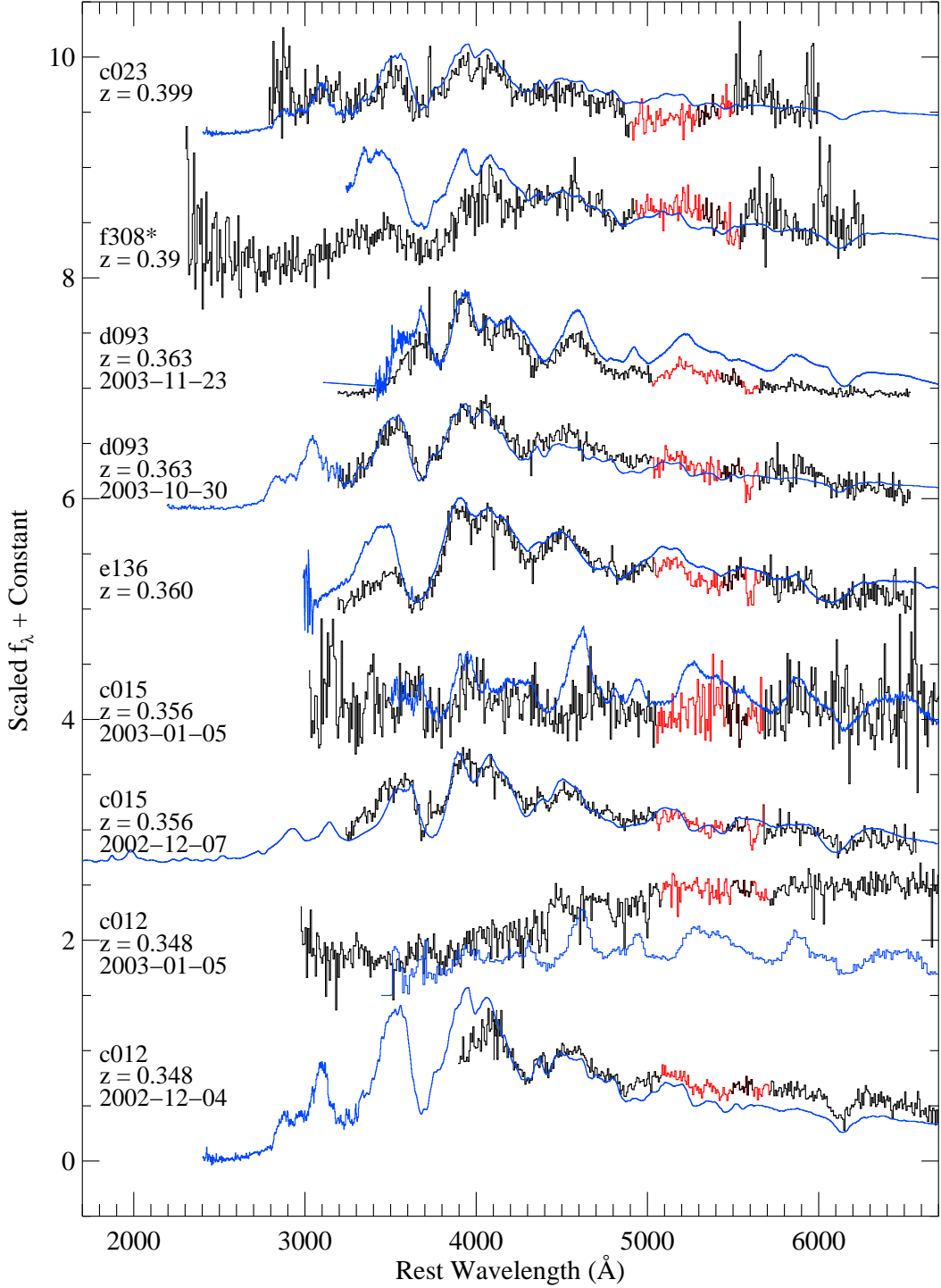}
\caption{Rest-wavelength spectra of ESSENCE SNe~Ia as in Figure 4a.
  The spectrum of f076 is a weighted average of the MMT and KI/LRIS
  spectra.  
  }
\end{figure}
\begin{figure}
\figurenum{4d}
\plotone{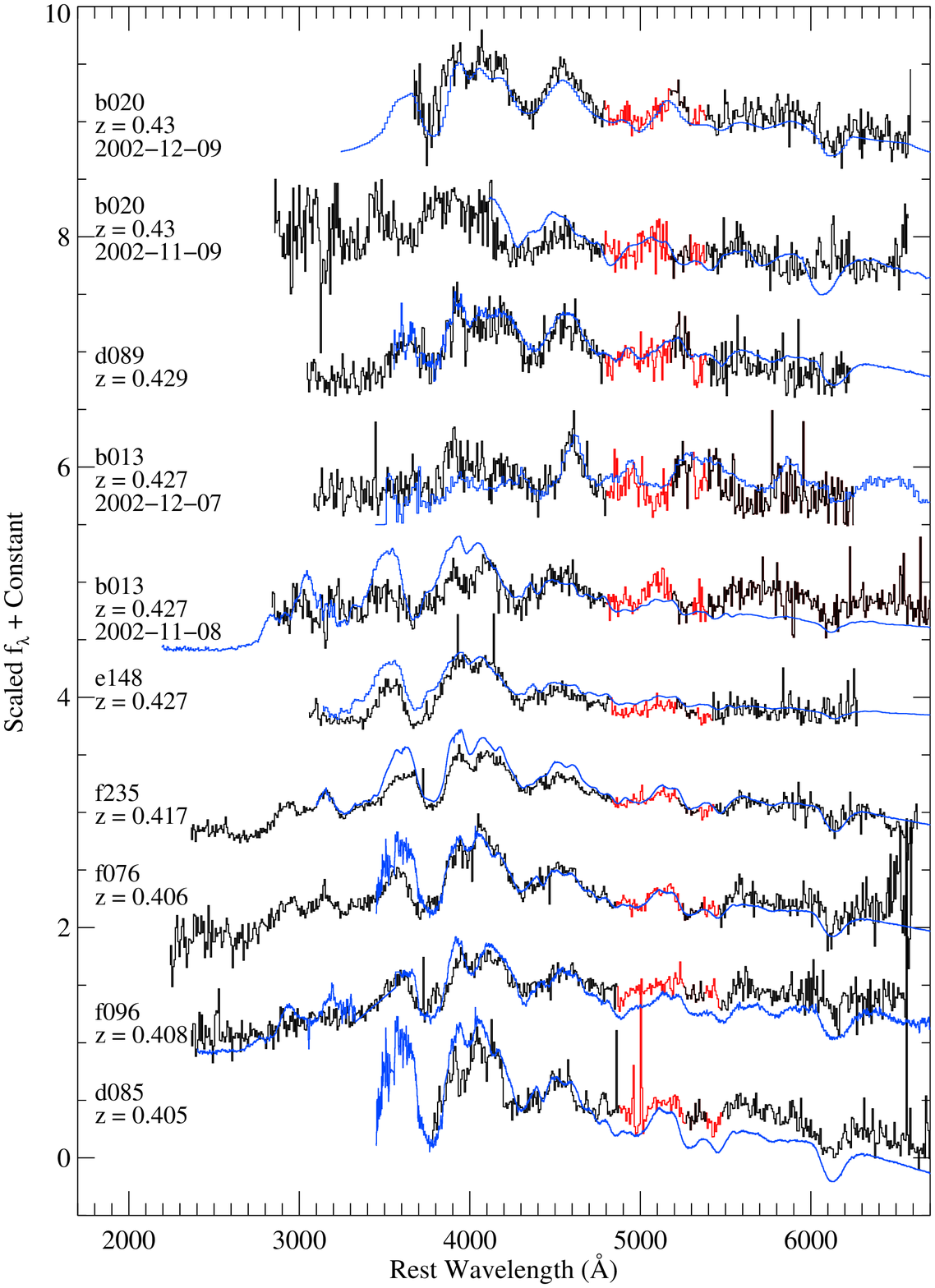}
\caption{Rest-wavelength spectra of ESSENCE SNe~Ia as in Figure 4a.
  }
\end{figure}

\begin{figure}
\figurenum{4e}
\plotone{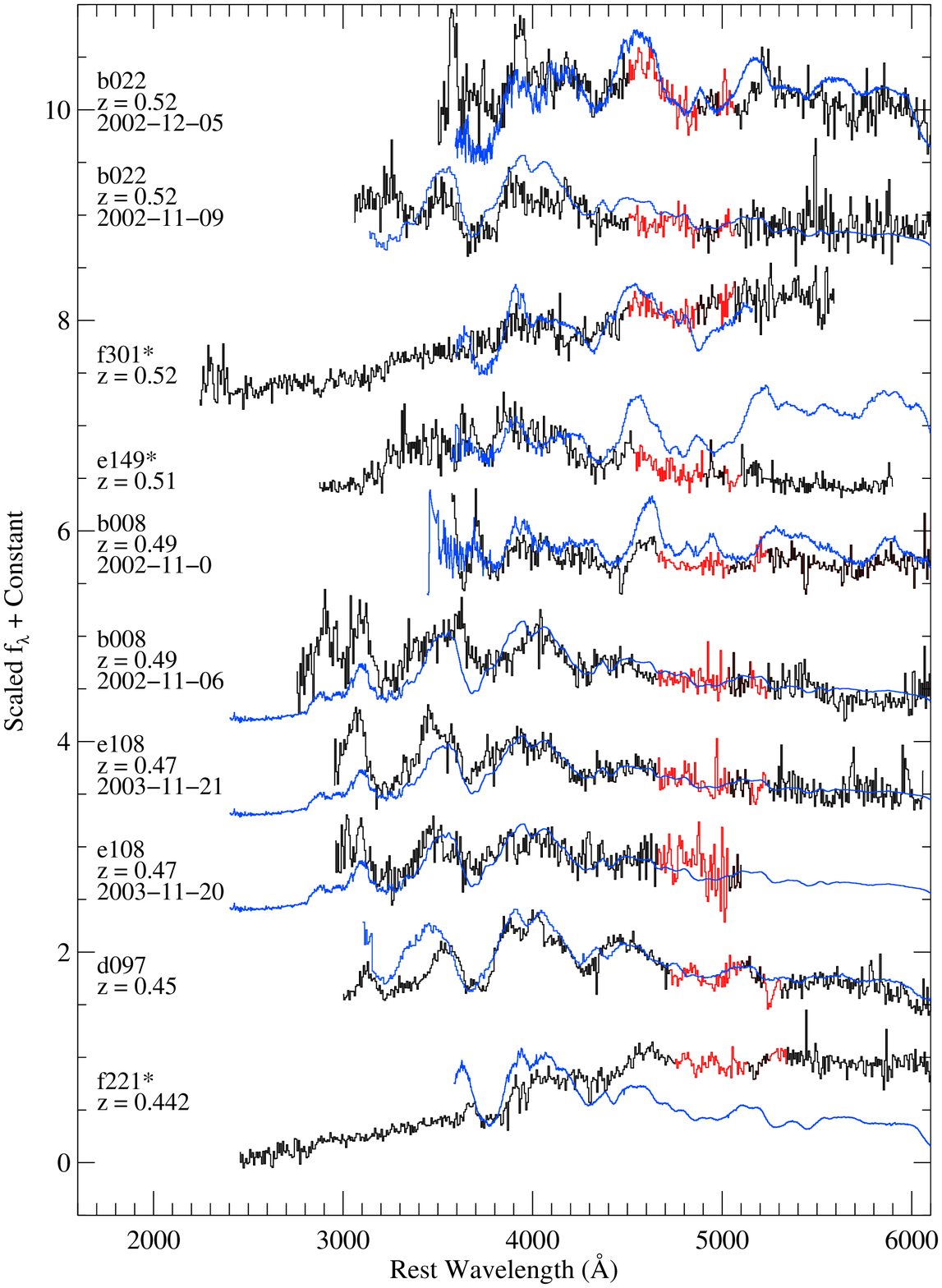}
\caption{Rest-wavelength spectra of ESSENCE SNe~Ia as in Figure 4a.
  }
\end{figure}
\begin{figure}
\figurenum{4f}
\plotone{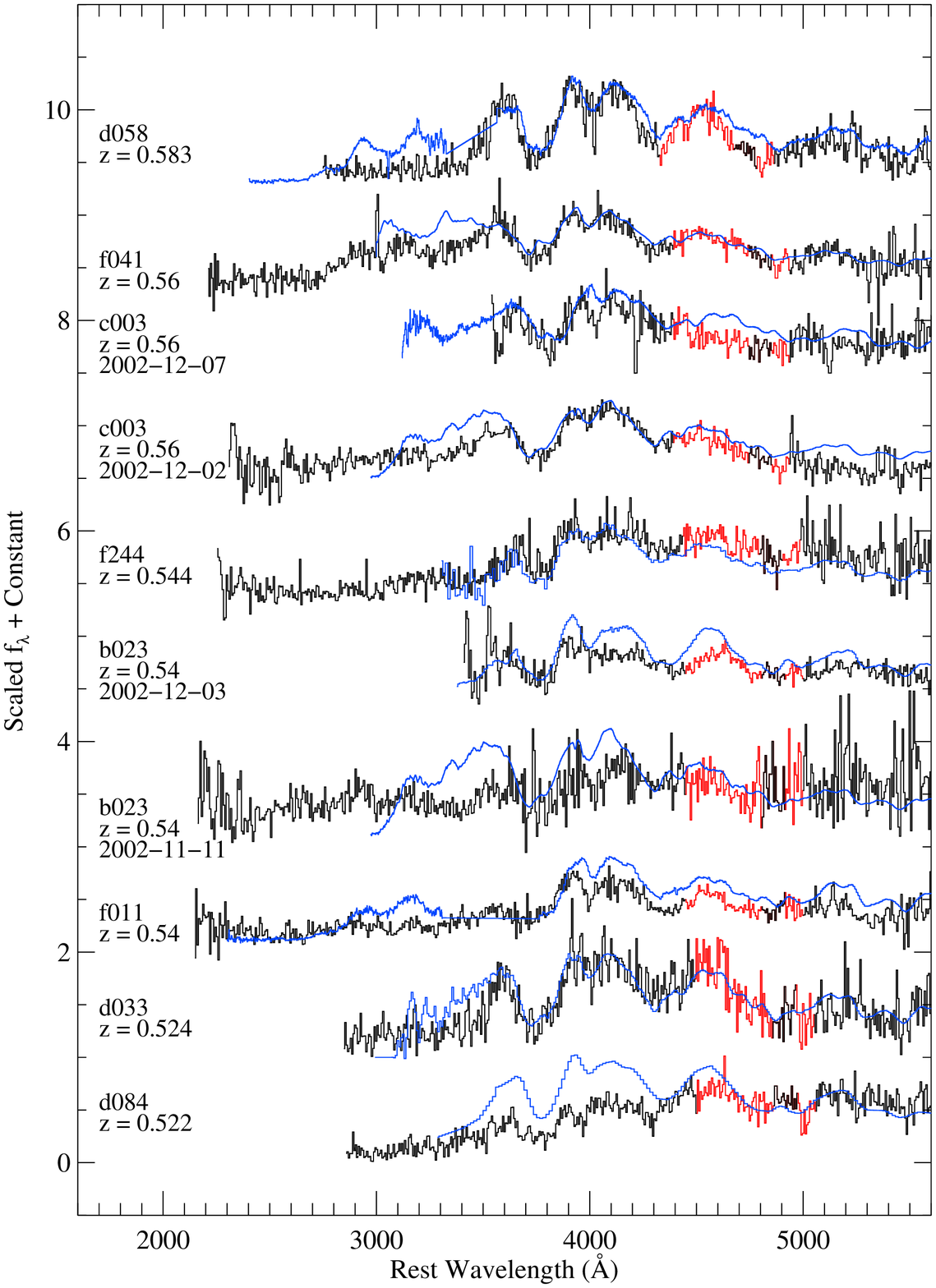}
\caption{Rest-wavelength spectra of ESSENCE SNe~Ia as in Figure 4a.
  }
\end{figure}
\begin{figure}
\figurenum{4g}
\plotone{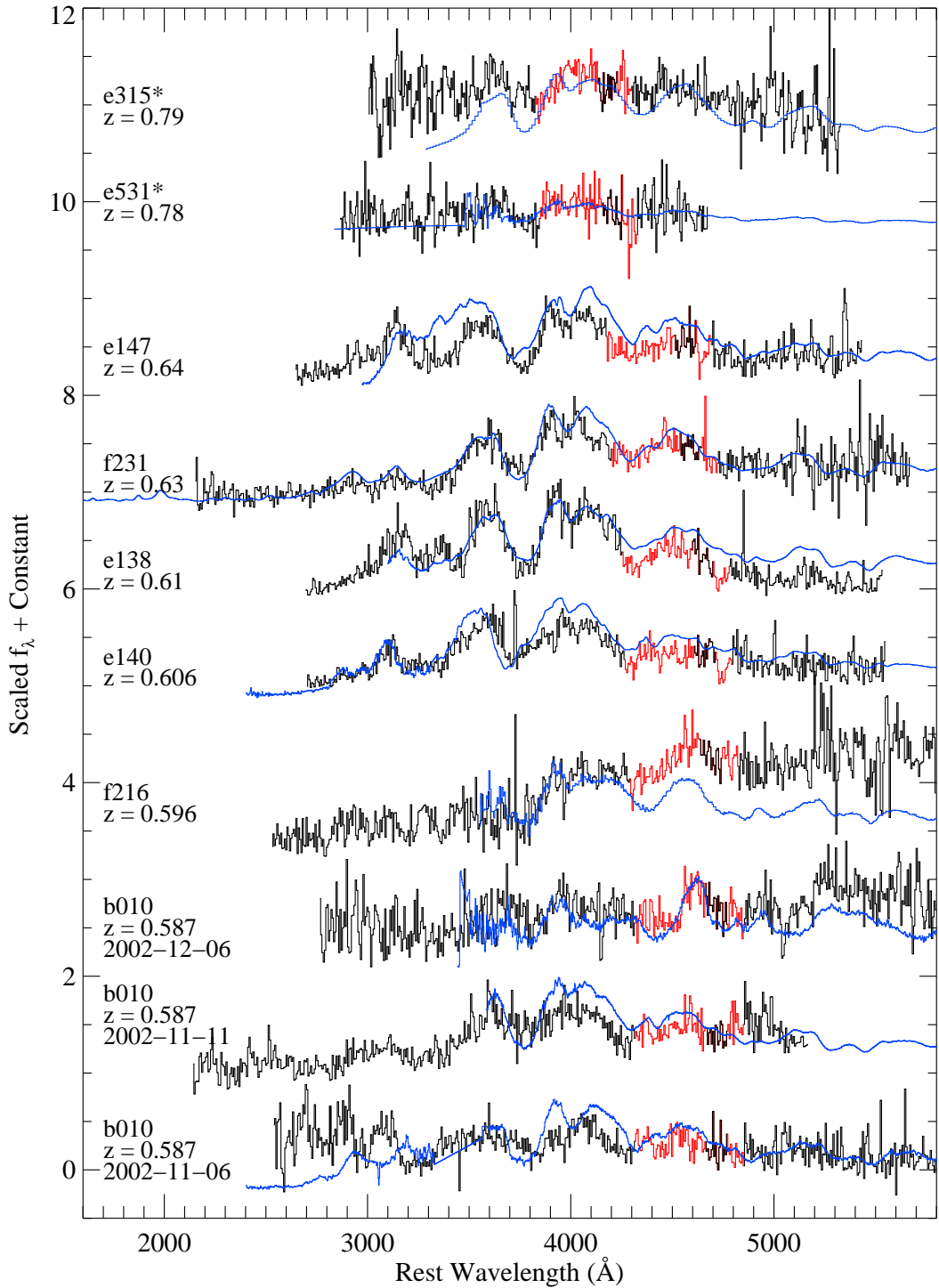}
\caption{Rest-wavelength spectra of ESSENCE SNe~Ia as in Figure 4a.  The GMOS and VLT spectra of b010 have been combined. }

\end{figure}

\begin{figure}
\addtocounter{figure}{1}
\plotone{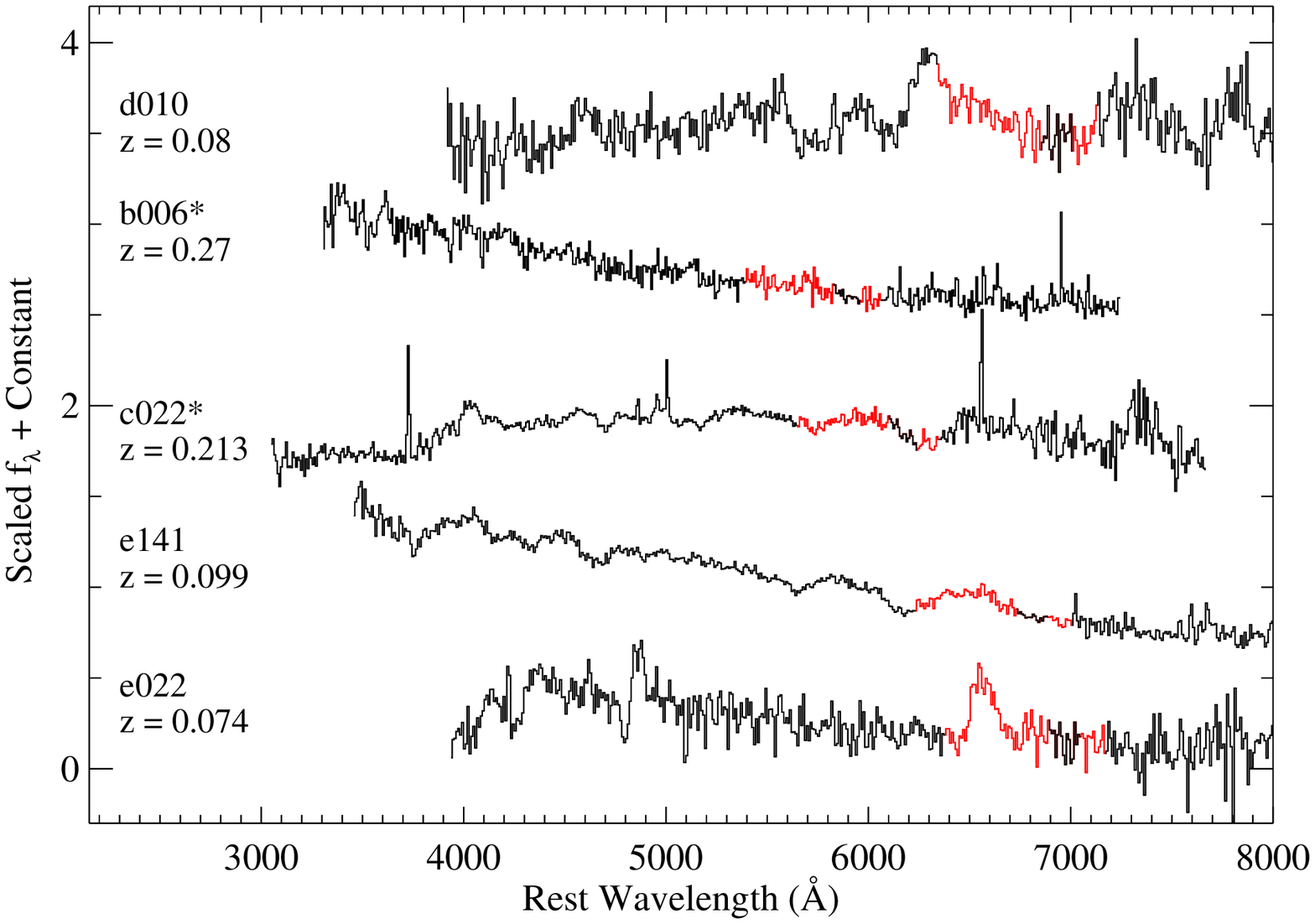}
\caption{Spectra of SNe~II and one SN~Ib/c from the first two years of
  the ESSENCE project.  Each spectrum is labeled with the ESSENCE
  identification number and the deduced redshift.  Spectra of
  uncertain SNe~II are indicated with an asterisk (*). The
  deredshifted regions of the spectra that are strongly affected by
  atmospheric absorption are shown in red.  The flux scale is
  $f_{\lambda}$ with arbitrary additive offsets between the spectra.
  The SN~Ib/c is d010 = SN~2003jp.\label{IIfig}}

\end{figure}

\begin{figure}
\figurenum{6a}
\plotone{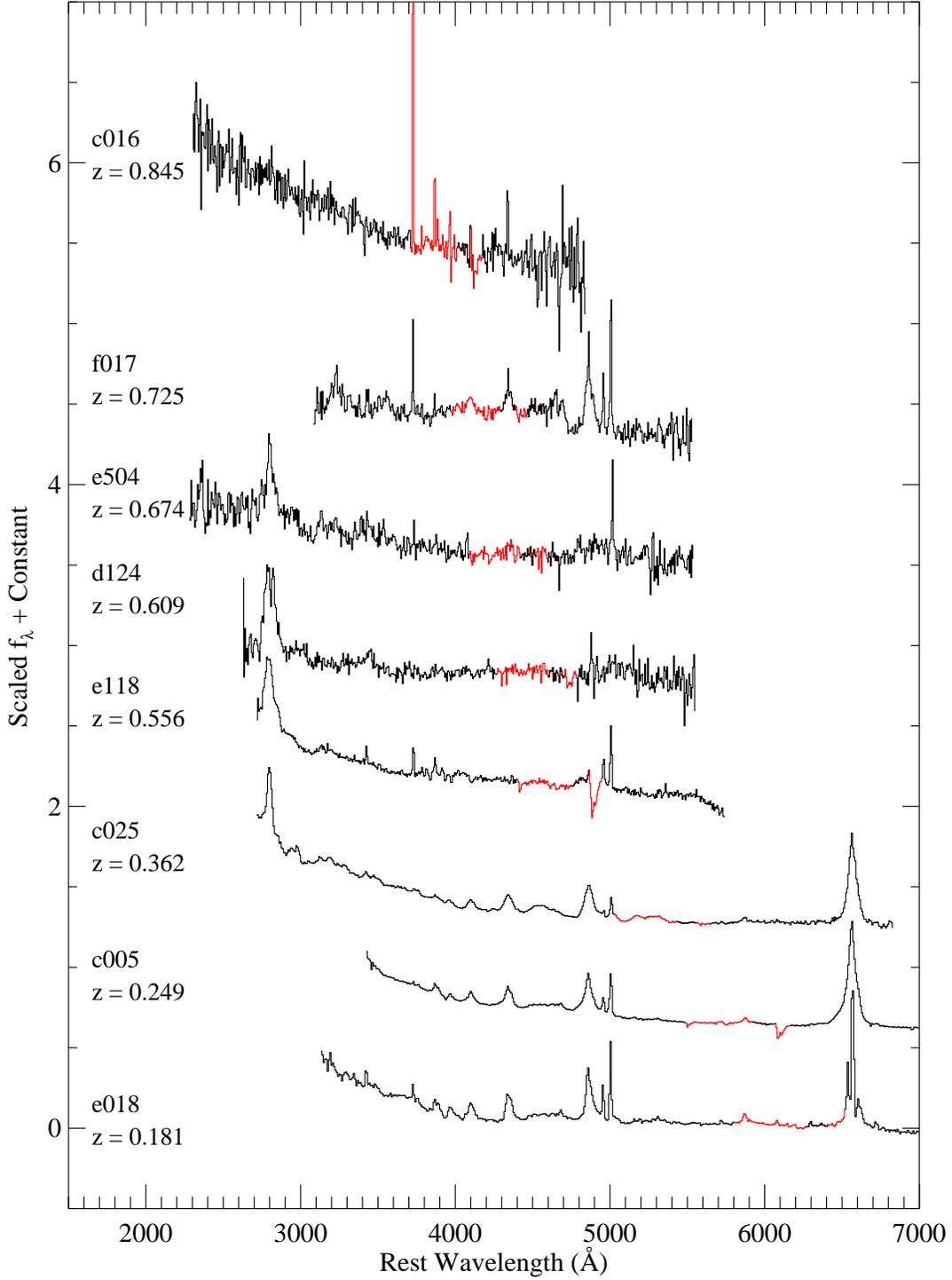}
\caption{Spectra of AGNs from the first two years of the ESSENCE
  project.  Each spectrum is labeled with the ESSENCE identification
  number and the deduced redshift.  The deredshifted regions of the
  spectra that are strongly affected by atmospheric absorption are
  shown in red.  The flux scale is $f_{\lambda}$ with arbitrary
  additive offsets between the spectra.\label{AGNfig}}

\end{figure}
\begin{figure}
\plotone{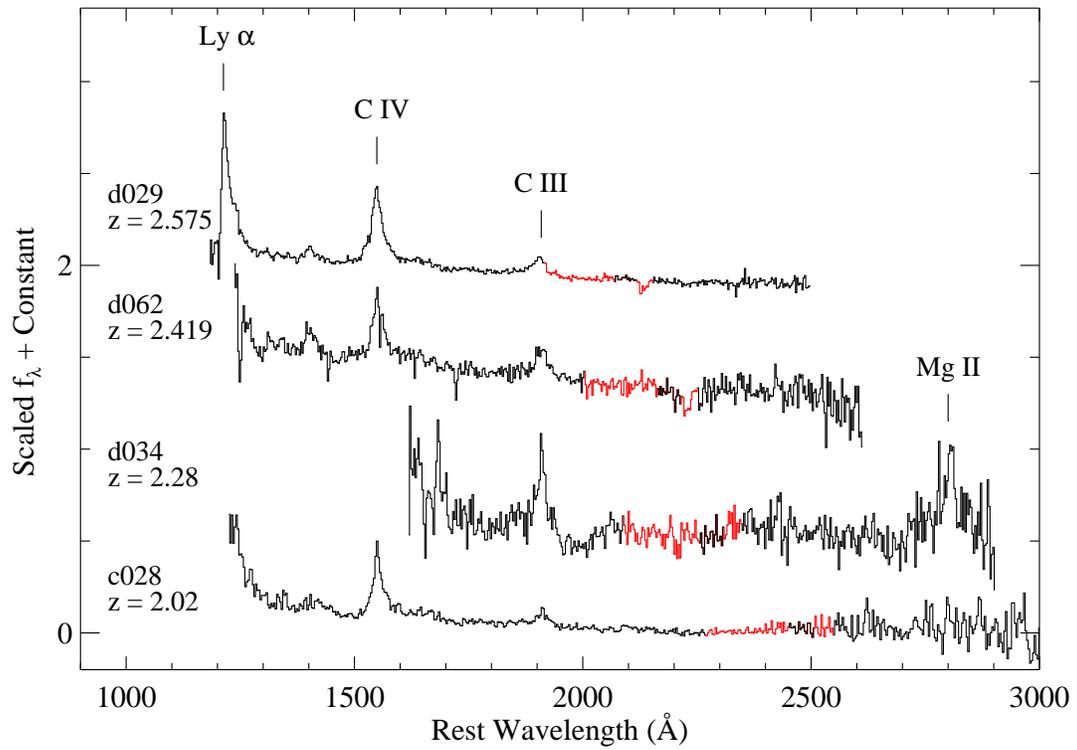}
\figurenum{6b}
\caption{Spectra of AGNs from the first
  two years of the ESSENCE project, as in Figure 6a.}

\end{figure}

\begin{figure}
\addtocounter{figure}{1}
\plotone{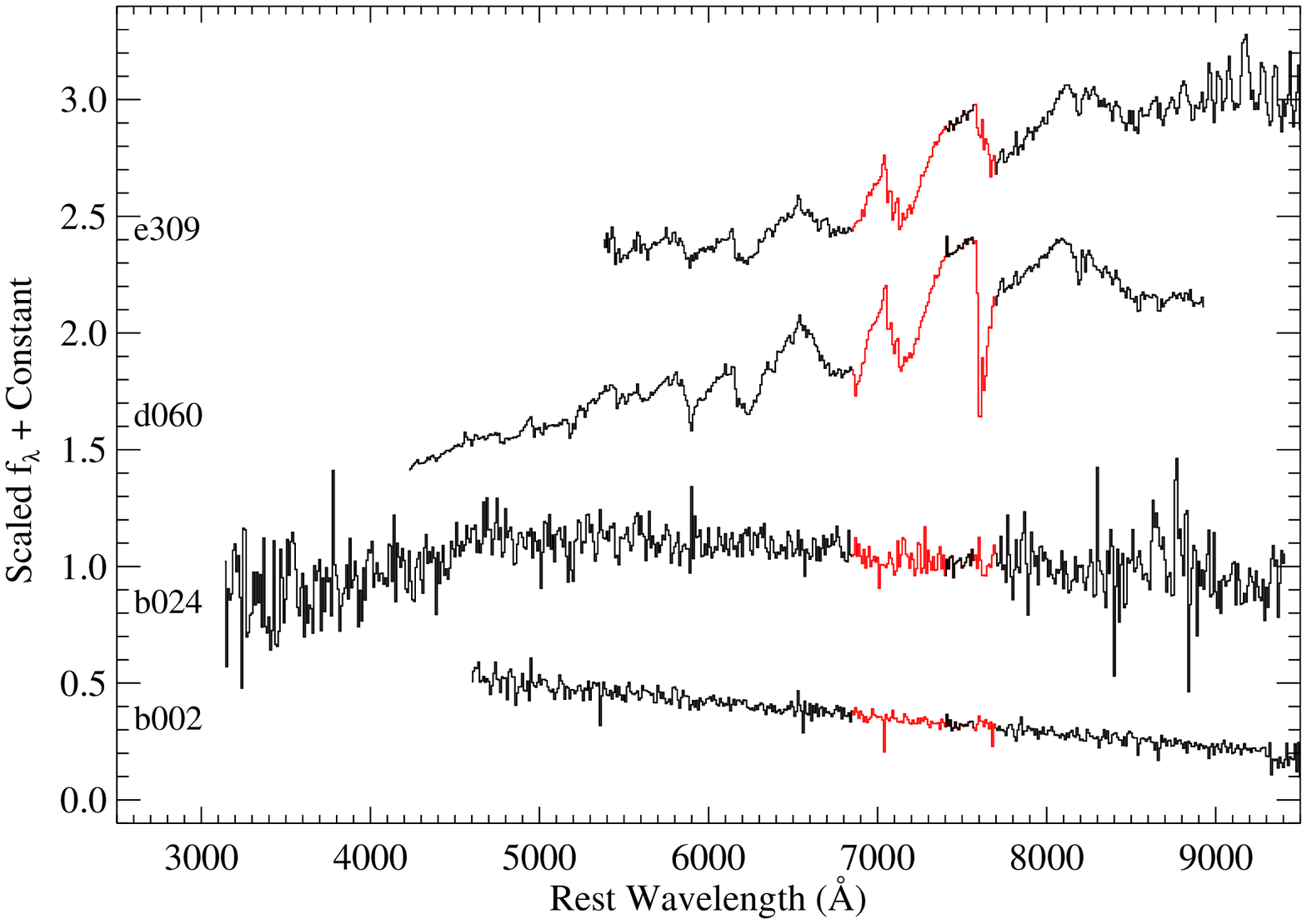}
\caption{Spectra of four stars from the first two years of the ESSENCE
  project.  Each spectrum is labeled with the ESSENCE identification
  number.  The deredshifted regions of the
  spectra that are strongly affected by atmospheric absorption are
  shown in red.  The flux scale is
  $f_{\lambda}$ with arbitrary additive offsets between the
  spectra.\label{starfig}}

\end{figure}
\begin{figure}
\figurenum{8a}
\plotone{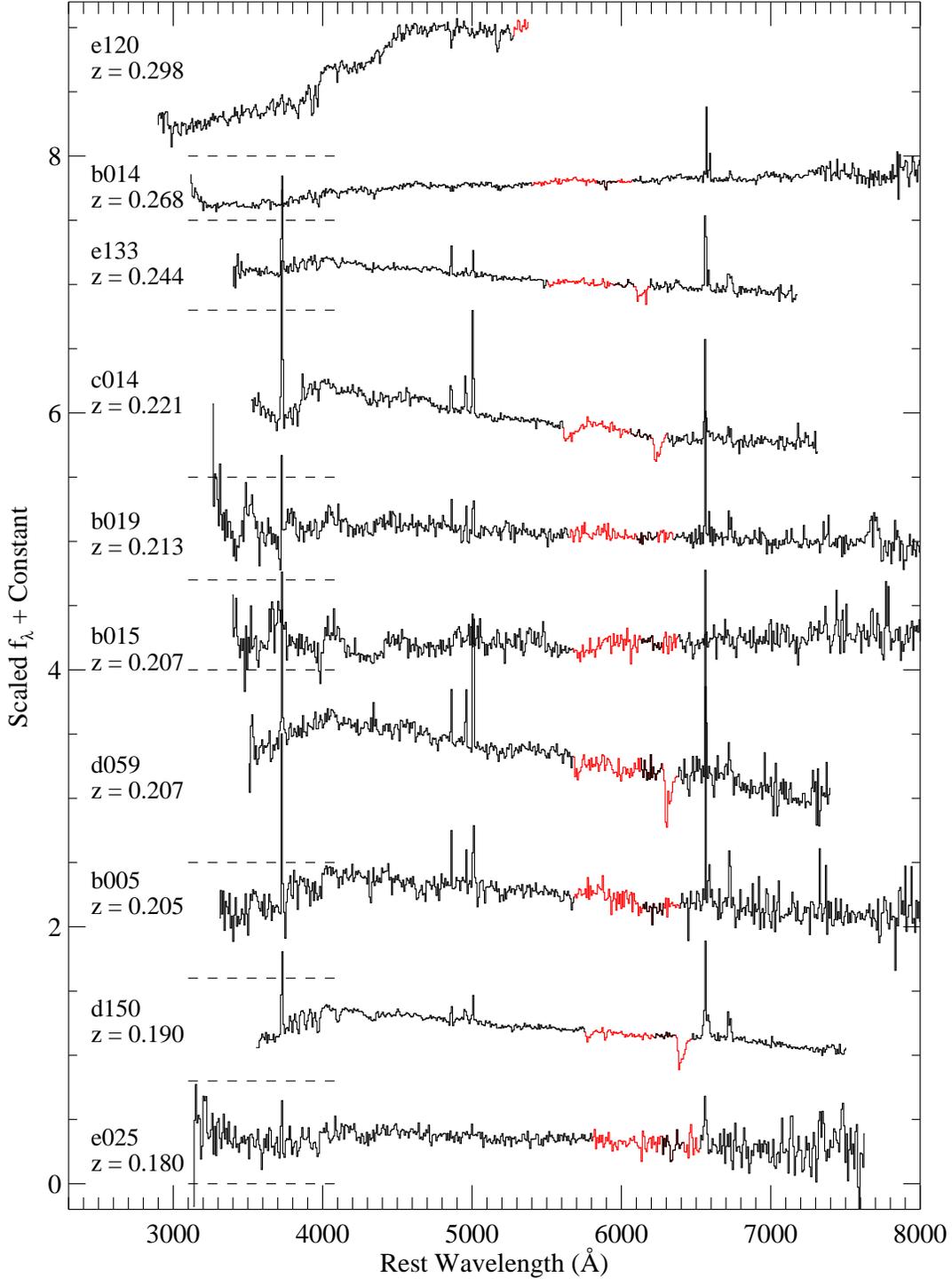}
\caption{Spectra of galaxies from the first two years of the ESSENCE
  project.  Each spectrum is labeled with the ESSENCE identification
  number and the deduced redshift.  The deredshifted regions of the
  spectra that are strongly affected by atmospheric absorption are
  shown in red.   The flux scale is $f_{\lambda}$ with arbitrary additive
  offsets between the spectra.  The zero-point of the flux scale for
  each spectrum is indicated (\emph{dashed line}).  For b005, the
  KII/ESI and MMT spectra have been combined.  For c014, the VLT and
  GMOS spectra have been combined.  \label{galfig}}

\end{figure}
\begin{figure}
\figurenum{8b}
\plotone{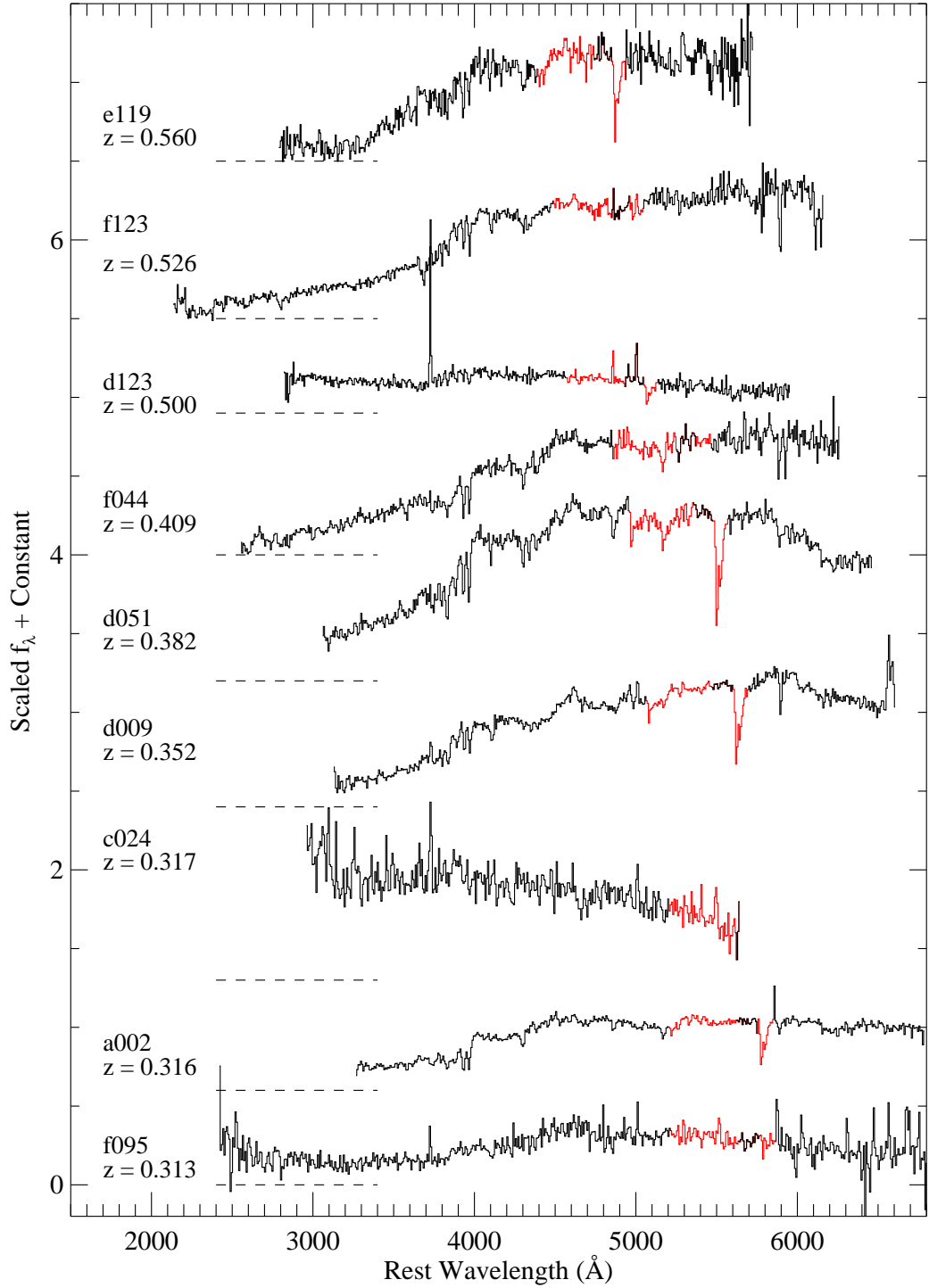}
\caption{Spectra of galaxies from the first two years of the ESSENCE
  project, as in Figure 8a.  For d009, the two VLT spectra have been combined.}

\end{figure}
\begin{figure}
\figurenum{9a}
\plotone{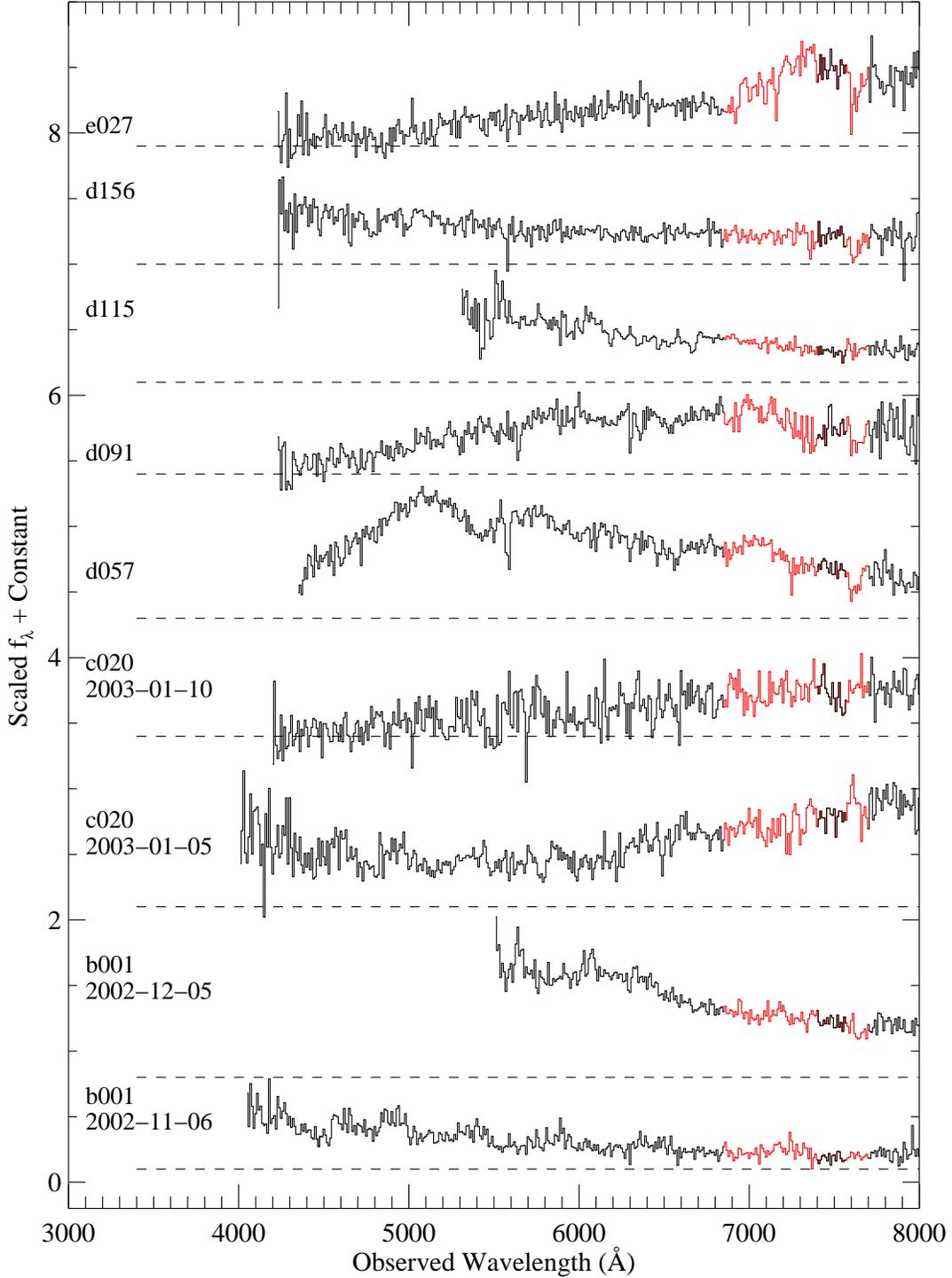}
\caption{Spectra of objects whose classification is uncertain from the
  first two years of the ESSENCE project.  Each spectrum is labeled
  with the ESSENCE identification number. The regions of the
  spectra that are strongly affected by atmospheric absorption are
  shown in red. 
  The flux scale is $f_{\lambda}$ with arbitrary additive offsets
  between the spectra.  The zero-point of the flux scale for each
  spectrum is indicated (\emph{dashed line}).}

\end{figure}
\begin{figure}
\figurenum{9b}
\plotone{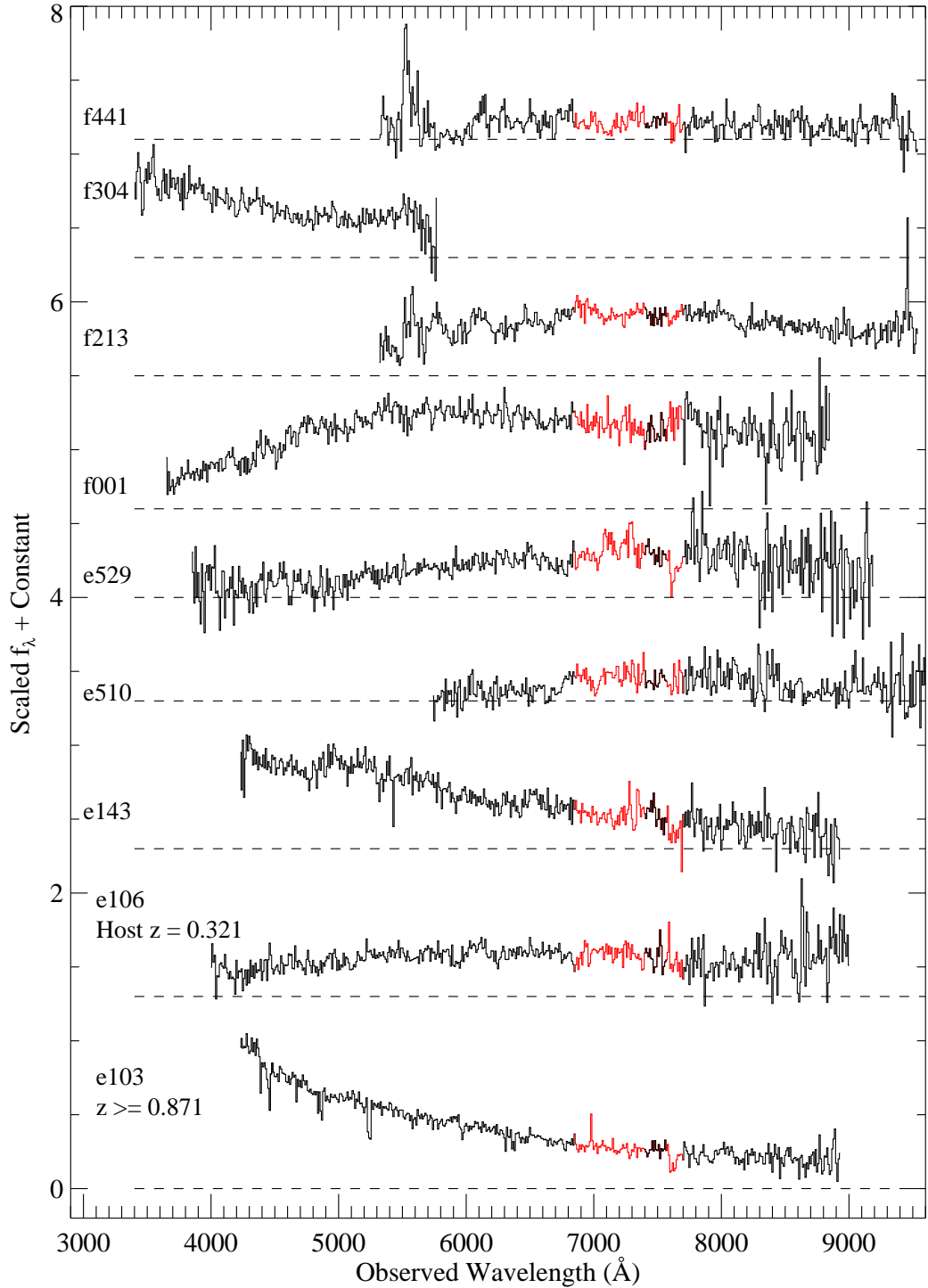}
\caption{Spectra of objects whose classification is uncertain from the
  first two years of the ESSENCE project, as in Figure 9a.  }

\end{figure}

No attempt has been made to remove host-galaxy contamination for any
object presented in Figures 4 and 5.  The amount of galaxy light is
significant for some objects (e.g., f221 on Figure 4d).  In addition,
no extinction corrections have been applied, either for Galactic
reddening or extinction in the host galaxy.  Given the Galactic
latitudes of the ESSENCE fields \citep{smith05} and our preference for
targets well separated from the host galaxy, the effects of extinction
are likely to be minimal \citep{blondin05, foley05}.

For each SN~Ia in Figure 4, the best match low-redshift comparison
spectrum as determined using SNID is included.  In general, the
high-redshift SNe~Ia look very similar to those at low redshift,
implying that there are no significant evolutionary effects.  Future
papers will deal in much greater detail with the comparison with
low-redshift SNe~Ia, as well as removal of galactic contamination and
the effects of extinction \citep{blondin05, foley05}.

While most of the high-redshift SNe~Ia appear to be normal, there are
some examples of peculiar SNe~Ia.  Both b004 (SN~2002iv; Figure 4a)
and d083 (SN~2003jn; Figure2b) show strong similarities with SN~1991T,
an overluminous Type Ia~SN \citep{filippenko92, phillips92}.  Given
the high rate of peculiar SNe~Ia at low redshift \citep{weidong01}, we
would expect to find such objects in a high-redshift sample.

We note that in Figure 4, there are several examples of high-redshift
SNe~Ia from the ESSENCE sample for which the low-redshift template
appears to be a poor match.  Examples include d086 (2003-10-27), b008
(2002-11-06), e149, and b022 (2002-12-05) .  The most likely
explanation for this is that the templates included in SNID do not
cover the complete range of possibilities, although problems with the
spectrum (e.g., poor S/N or sky subtraction) may also play a role.
Rather than perform a non-objective search for the best match, we
leave these as examples of the current limitations in SNID.  We plan
to expand the SNID templates to eliminate such occurrences in the
future.

Two unusual cases in the sample of SNe~Ia spectra are e315 (SN~2003ku)
and b004 (SN~2002iv).  They are the only SNe for which the redshift
determination in SNID was ambiguous.  For e315, if we assume that it
is a Ia, then the best fit redshift is 0.79, but a fit at a redshift
of 0.41 is only marginally worse.  All other SNe~Ia spectra in the
ESSENCE sample had redshifts determined by SNID that were unambiguous.
If we rely on the spectrum alone, the SNID result of $z=0.79$ is what
we would choose, and so we report it in this paper.  Analysis
presented by \citet{krisciunas05} shows that neither of the redshifts
suggested by SNID gives a very satisfactory fit to the photometry.  In
the case of b004, the SNID result is $z=0.39$, while the fit when
$z=0.231$, known from galaxy features, is almost as good.  As b004 is
similar to SN~1991T (see above), the lack of good templates in SNID
may be the source of this discrepancy.  Correlation of b004 with
SN~1991T templates yields a redshift of $z=0.22$, close to the value
derived from the host galaxy.  This highlights some of the perils of
identifying optical transients with low S/N spectra.  Sometimes the
spectrum alone is not enough.  A consideration of all the information
(spectrum, light curve, host galaxy, etc.) is necessary to draw the
appropriate conclusion.

Among the unknown spectra (Figure 9a), there are three spectra that
require some discussion.  For e106, the redshift is known because the
host galaxy was also observed.  The emission line appearing at the
observed wavelength of 9457 \AA\ in f213 is real.  If this is
H$\alpha$ then $z=0.44$; if it is [\ion{O}{3}] $\lambda$5007, then
$z=0.89$.  There is an apparent doublet absorption line in e103 at an
observed wavelength of 5240 \AA.  We interpret this as \ion{Mg}{2}
$\lambda$2800 at a redshift of 0.871, implying that this object has a
redshift at least that high.  It is likely to be a high-redshift AGN,
but we do not have enough information to move it out of the unknown
category.

\section{Conclusions}

We have presented optical spectroscopy of the targets selected for
follow-up observations from the first two years of the ESSENCE
project.  As the target selection process has improved, we have
increased our yield of SNe~Ia that are needed for the primary purpose
of the ESSENCE project---measuring luminosity distances to $\sim$200
SNe~Ia over the redshift range $(0.2 \lesssim z \lesssim 0.8)$.  The
SNe~Ia show strong similarities with low-redshift SNe~Ia, implying
that there are no significant evolutionary changes in the nature of
Type Ia~SNe and that our methods for identifying objects have been
successful.  This is also shown by the concordance of redshifts
derived from SN spectra and those found from the host galaxy itself.
Over the next three years, ESSENCE will continue to discover
high-redshift SNe~Ia.  With enough spectroscopic telescope time, we
plan to be even more successful in correctly identifying Type Ia~SNe
than we have been during the first two years.

All spectra presented in this paper will be made publicly available
upon publication.

\acknowledgments

We would like to thank the staffs of the Paranal, Gemini, Keck, Las
Campanas, MMT, F.~L. Whipple, and Cerro Tololo Inter-American
Observatories for their extensive assistance and support during this
project.  We would also like to thank Warren Brown and Craig Heinke
for assistance with the MMT observations.  This work is supported
primarily by NSF grants AST-0206329 and AST-0443378.  In addition,
A.V.F.'s group at UC Berkeley acknowledges NSF grant AST-0307894.
C.W.S thanks the McDonnell Foundation and Harvard University for their
support.  A.C. acknowledges the support of CONICYT (Chile) through
FONDECYT grants 1000524 and 7000524.

\begin{deluxetable}{rrr|r}
%\tabletypesize{\scriptsize}
\tablenum{1}
\tablewidth{0pt}
\tablecaption{ESSENCE SPECTROSCOPY RESULTS: THE FIRST TWO YEARS\label{resulttable}}
\tablehead{\colhead{Type\tablenotemark{a}}  &
\colhead{Year 1} &
\colhead{Year 2} &
\colhead{Total}
}

\startdata
Ia  &  15  & 31  & 46  \\
Ia? &   0  &  6  &  6  \\
II  &   2  &  2  &  4  \\
Ib/c&   0  &  1  &  1  \\
AGN &   4  &  8  & 12  \\
Gal &   7  & 12  & 19  \\
star&   2  &  2  &  4  \\
N.S.&   5  &  5  & 10 \\
Unk.&  2   & 14  & 16  \\
N.A.&  13  & 43  & 54  \\ \tableline
Total & 50 & 124 & 174 \\
\enddata

\tablenotetext{a}{Our best guess as to classification of the object.
  Ia? indicates a lack of certainty in the identification as a
  SN~Ia.  N.S. indicates that the telescope was pointed to the object,
  but no spectrum was obtained.  Unk. represents objects for which we
  have spectra, but are uncertain as to their classification.
  N.A. indicates that a transient was found in the ESSENCE search, but
  no attempt to take a spectrum was made, either because it was a poor
  target or there were not enough spectroscopic resources available.}

\end{deluxetable}                                
\clearpage

\begin{deluxetable}{llllllllllll}
\tabletypesize{\scriptsize}
\rotate
\tablenum{2}
\tablewidth{0pt}
\tablecaption{ESSENCE SPECTROSCOPIC TARGETS: THE FIRST TWO YEARS\label{sntable}}
\tablehead{\colhead{ESSENCE ID\tablenotemark{a}} & 
\colhead{IAUC ID\tablenotemark{b}} &
\colhead{UT Date\tablenotemark{c}} &
\colhead{Telescope} &
\colhead{Type\tablenotemark{d}}  &
\colhead{$z$\tablenotemark{e}} &
\colhead{$z$\tablenotemark{f}} &
\colhead{Epoch\tablenotemark{g}} &
\colhead{Template\tablenotemark{h}} &
\colhead{Grade\tablenotemark{i}} &
\colhead{Disc.\tablenotemark{j}} & 
\colhead{Exp.} \\
\colhead{} &
\colhead{} &
\colhead{} &
\colhead{} &
\colhead{} &
\colhead{ (Gal)} &
\colhead{ (SNID)} &
\colhead{(SNID)} &
\colhead{} &
\colhead{} & 
\colhead{Mag.} & 
\colhead{(s)}
}

\startdata
a002.wxc1\_04  & \nodata &  2002-12-06.03  & VLT       & Gal    & 0.316 &\nodata&\nodata &  \nodata    &\nodata  &\nodata     & 900                \\
b001.wxc1\_14  &\nodata  &  2002-11-06.27  & KII/ESI   & Unk    &\nodata&\nodata& \nodata&\nodata& \nodata & 23.6       & 1600           \\
b001.wxc1\_14  &\nodata  &  2002-11-11.32  & KI/LRIS   & Unk    &\nodata&\nodata&\nodata&\nodata& \nodata & 23.6       & 1800           \\
b001.wxc1\_14  &\nodata  &  2002-12-05.26  & GMOS      & Unk    &\nodata&\nodata&\nodata & \nodata     &\nodata  & 23.6       & 2x1800         \\
b002.wxh1\_01  &\nodata  &  2002-11-01.44  & KII/ESI   & star   &\nodata&\nodata&\nodata & \nodata     &\nodata  & \nodata    & 900            \\
b003.wxh1\_14  & 2002iu  &  2002-11-01.43  & KII/ESI   & Ia     &\nodata& 0.11  &  2     & 94S  & good    & 18.9       & 600            \\
b004.wxt2\_06  & 2002iv  &  2002-11-02.45  & KII/ESI   & Ia     & 0.231 & 0.39  & -4     & 95ac & good    & 20.9       & 1200           \\
b005.wxd1\_11  & 2002iw  &  2002-11-03.16  & MMT       & Gal    & 0.205 &\nodata&\nodata &\nodata      &\nodata  & 21.8       & 3x1800         \\
b005.wxd1\_11  & 2002iw  &  2002-11-06.32  & KII/ESI   & Gal    & 0.205 &\nodata&\nodata &\nodata      &\nodata  & 21.8       & 1800           \\
b006.wxb1\_16  & 2002ix  &  2002-11-03.10  & MMT       & N.S.   &\nodata&\nodata&\nodata &\nodata      &\nodata  & 22.2       &\nodata         \\
b006.wxb1\_16  & 2002ix  &  2002-11-06.24  & KII/ESI   & II?    &\nodata&\nodata&\nodata &\nodata      &\nodata  & 22.2       & 1800           \\
b008.wxc1\_05  & 2002jq  &  2002-11-06.29  & KII/ESI   & Ia     &\nodata& 0.49  & -8     & 90N  & poor    & 21.9       & 1800           \\
b008.wxc1\_05  & 2002jq  &  2002-12-04.27  & GMOS      & Ia     &\nodata& 0.49  & +13    & 89B  & good    & 21.9     &   4x1800         \\
b010.wxv2\_07  & 2002iy  &  2002-11-06.39  & KII/ESI   & Ia     &\nodata& 0.59  & -1     & 92A  & poor    & 21.3       & 1800           \\
b010.wxv2\_07  & 2002iy  &  2002-11-11.45  & KI/LRIS   & Ia     &\nodata& 0.59  & +2     & 94ae & poor    & 21.3       & 1800           \\
b010.wxv2\_07  & 2002iy  &  2002-12-06.35  & GMOS      & Ia     &\nodata& 0.59 & +13    & 89B  & good    & 21.3       & 5x1800          \\
b010.wxv2\_07  & 2002iy  &  2002-12-07.22  & VLT       & Ia     & 0.587 & 0.59  & +17    & 95al & good    & 21.3       & 2x1800         \\
b013.wxv2\_10  & 2002iz  &  2002-11-06.45  & KII/ESI   & Ia     & 0.427 & 0.42  & -6     & 90N  & good    & 22.1       & 1800           \\
b013.wxv2\_10  & 2002iz  &  2002-12-06.07  & VLT       & Ia     & 0.428 & 0.43  & +14    & 89B  & good    & 22.1       & 1800           \\
b014.wxv2\_15  &\nodata  &  2002-11-06.41  & KII/ESI   & Gal    & 0.268 &\nodata&\nodata & \nodata     &\nodata  & 22.8       & 1800           \\
b015.wcx1\_09  &\nodata  &  2002-11-06.34  & KII/ESI   & Gal    & 0.207 &\nodata&\nodata & \nodata     &\nodata  & \nodata    & 1800           \\
b016.wxb1\_15  & 2002ja  &  2002-11-09.33  & KII/ESI   & Ia     &\nodata& 0.33  & +2     & 94ae & good    & 22.2       & 1200           \\
b017.wxb1\_06  & 2002jb  &  2002-11-09.31  & KII/ESI   & Ia     &\nodata& 0.25  & +2     & 94ae & good    & 21.4       & 1200           \\
b019.wxd1\_04  &\nodata  &  2002-11-09.35  & KII/ESI   & Gal    & 0.213 &\nodata&\nodata & \nodata     &\nodata  & 22.2       & 1200           \\
b020.wye2\_01  & 2002jr  &  2002-11-09.42  & KII/ESI   & Ia     &\nodata& 0.43  & -9     & 91M  & poor    & 22.7       & 1800           \\
b020.wye2\_01  & 2002jr  &  2002-12-09.39  & GMOS      & Ia     &\nodata& 0.43  & +7     & 72E  & good    & 22.7       & 2x1800         \\
b022.wyc3\_03  & 2002jc  &  2002-11-09.45  & KII/ESI   & Ia     &\nodata& 0.52  & -7     & 90N  & poor    & 23.1       & 1800           \\
b022.wyc3\_03  & 2002jc  &  2002-12-05.39  & GMOS      & Ia     &\nodata& 0.52  & +4     & 96Z  & good    & 23.1       & 4x1800         \\
b023.wxu2\_09  & 2002js  &  2002-11-11.57  & KI/LRIS   & Ia     &\nodata& 0.54  & -7     & 89B  & poor    & 22.9       & 2100           \\
b023.wxu2\_09  & 2002js  &  2002-12-03.41  & GMOS      & Ia     &\nodata& 0.54  & +6     & 92A  & good    & 22.9       &  4x1800        \\
b024.wxc1\_16  &\nodata  &  2002-11-11.27  & KI/LRIS   & VStar  &\nodata&\nodata&\nodata &  \nodata    &\nodata  & 21.5       & 1800           \\
b025.wxa1\_05  &\nodata  &  2002-11-11.29  & KI/LRIS   & N.S.   &\nodata&\nodata&\nodata & \nodata     &\nodata  & 22.2       & 1800           \\
b026.wxk1\_05  &\nodata  &  2002-11-11     & KI/LRIS   & N.S.   &\nodata&\nodata      &\nodata  & \nodata     &\nodata  & 22.2       &       \nodata  \\
b027.wxm1\_16 & 2002jd  &  2002-11-11.45  & KI/LRIS   & Ia     & \nodata  & 0.32  & +0     & 81B  & good    & 22.0       & 3600,1800  {\tablenotemark{k}}      \\
b027.wxm1\_16  & 2002jd  &  2002-12-06.07  & VLT       & Ia     &\nodata & 0.32 & +13    & 89B  & good    & 22.0       & 1800           \\
b027.wxm1\_16  & 2002jd  &  2002-12-09.27  & GMOS      & Ia     &\nodata & 0.32 & +12    & 92G  & good    & 22.0       & 4x1800         \\
c002.wxp1\_14  &\nodata  &  2002-12-01     & Clay      & N.S.   &\nodata&\nodata&\nodata & \nodata     &\nodata  & 22.4       &\nodata         \\
c002.wxp1\_14  &\nodata  &  2002-12-03     & GMOS      & N.S.   &\nodata&\nodata&\nodata & \nodata     &\nodata  & 22.4       &\nodata         \\
c003.wxh1\_15  & 2002jt  &  2002-12-02.13  & Clay      & Ia     &\nodata& 0.56  & -7     & 89B  & good    & 22.6       & 2x1800         \\
c003.wxh1\_15  & 2002jt  &  2002-12-07.29  & GMOS      & Ia     &\nodata& 0.56  & +0     & 94S  & good    & 22.6       & 3x1800         \\
c005.wxb1\_10  & \nodata &  2002-12-06.05  & VLT       & AGN    & 0.249 &\nodata&\nodata & \nodata     &\nodata  & \nodata    & 900            \\
c012.wxu2\_16{\tablenotemark{l}}  & 2002ju  &  2002-12-03.15  & Clay      & Ia     & 0.348 & 0.35  & -8     & 90N  & good & 21.6       & 2x1800         \\
c012.wxu2\_16 {\tablenotemark{l}}  & 2002ju  &  2002-12-04.43  & GMOS      & Ia     & 0.348 & 0.35  & -8     & 90N  & good    & 21.6       & 3x1200         \\
c012.wxu2\_16  & 2002ju  &  2003-01-05.33  & KII/ESI   & Ia     & 0.348 & 0.35  & +14    & 89B  & poor    & 21.6       & 1800           \\
c013.wxm1\_13  & \nodata &  2002-12-05     & GMOS      & N.S.   &\nodata&\nodata&\nodata & \nodata     &\nodata  &\nodata     &       \nodata  \\
c014.wyb3\_03  & 2002jv  &  2002-12-07.12  & VLT       & Gal    & 0.221 &\nodata&\nodata & \nodata      &\nodata  & 22.6       & 2x1800         \\
c014.wyb3\_03  & 2002jv  &  2003-01-04.28  & GMOS      & Gal    & 0.221 &\nodata&\nodata &  \nodata    &\nodata  & 22.6       & 4x1800         \\
c015.wxv2\_02  & 2002jw  &  2002-12-07.18  & VLT       & Ia     & 0.357 & 0.35  & +0     & 81B  & good    & 22.8       & 2x1800         \\
c015.wxv2\_02  & 2002jw  &  2003-01-05.37  & KII/ESI   & Ia     & 0.356 & 0.38  & +12    & 89B & poor    & 22.8       & 2400           \\
c016.wxm1\_04  & \nodata &  2002-12-07.07  & VLT       & AGN    & 0.845 &\nodata&\nodata & \nodata     &\nodata  & 23.5       & 2x1800         \\
c020.wxt2\_15  &\nodata  &  2003-01-05.30  & KII/ESI   & Unk    & 0.650 &\nodata&\nodata & \nodata     &\nodata  & 23.3       & 2x900          \\
c020.wxt2\_15  & \nodata &  2003-01-10.11  & VLT       & Unk    &\nodata&\nodata&\nodata & \nodata     &\nodata  & 23.3       & 2x1800         \\
c022.wxu2\_15  & \nodata &  2003-01-04.09  & Clay      & II?    & 0.213 &\nodata&\nodata & \nodata     &\nodata  & \nodata    & 2x1200         \\
c023.wxm1\_15  & \nodata &  2003-01-03.07  & Clay      & Ia     & 0.399 & 0.42  & -8     & 90N  & good    & \nodata    & 2x1200         \\
c024.wxv2\_05  & \nodata &  2003-01-03.12  & Clay      & Gal    & 0.317 &\nodata&\nodata & \nodata     &\nodata  & \nodata    & 1800  \\
c025.wxb1\_14  & \nodata &  2003-01-04.05  & Clay      & AGN    & 0.362 &\nodata&\nodata &\nodata      &\nodata  & \nodata    & 600            \\
c028.wxu2\_16  & \nodata &  2003-01-04.13  & Clay      & AGN    & 2.02  &\nodata&\nodata &\nodata      &\nodata  & \nodata    & 1800           \\
d009.waa6\_16 {\tablenotemark{m}} & \nodata &  2003-10-29.11  & VLT  & Gal  & 0.352 & \nodata  &\nodata  & \nodata     &\nodata  & \nodata    & 1800         \\
d106.waa6\_16 {\tablenotemark{m}} & \nodata &  2003-10-31.01  & VLT  & Gal  & 0.353 & \nodata  &\nodata  & \nodata     &\nodata  & \nodata    & 1800         \\
d010.waa6\_16  & 2003jp  &  2003-10-30.03  & VLT       & Ib/c   &\nodata& 0.08  & +35    & 87M{\tablenotemark{n}} & poor    & 21.6       & 2x1800         \\
d029.waa6\_13  & \nodata &  2003-10-29.03  & VLT       & AGN    & 2.575 &\nodata&\nodata & \nodata     &\nodata  & 21.6       & 2x1800         \\
d033.waa6\_10  & 2003jo  &  2003-10-29.09  & VLT       & Ia     & 0.524 & 0.53  & -1     & 89B  & good    & 20.9       & 2x1800         \\
d033.waa6\_10  & 2003jo  &  2003-11-23.05  & VLT       & N.S.   & 0.524 &\nodata&\nodata & \nodata     &\nodata  & 20.9       & 2x1800         \\
d034.waa7\_10  & \nodata &  2003-10-28.29   & GMOS      & AGN    & 2.28  &\nodata      &\nodata &  \nodata    &\nodata  & 21.4       & 2x1200         \\
d051.wcc8\_2   & \nodata &  2003-10-30.22  & VLT       & Gal    & 0.382 &\nodata&\nodata &   \nodata   &\nodata  &\nodata     & 1800           \\
d057.wbb6\_3   & 2003jk  &  2003-10-30.15  & VLT       & Unk    &\nodata & \nodata     &\nodata &  \nodata    &\nodata  & 20.9       & 2x1800         \\
d058.wbb6\_3   & 2003jj  &  2003-10-31.07  & VLT       & Ia     & 0.583 & 0.58  & -1     & 92A  & good    & 23.1       & 2x1800         \\
d059.wcc5\_3   & \nodata &  2003-10-29.17  & VLT       & Gal    & 0.207 &\nodata&\nodata & \nodata     &\nodata  & 19.2       & 2x1800         \\
d060.wcc7\_3   & \nodata &  2003-10-30.35  & VLT       & M-star &\nodata&\nodata&\nodata & \nodata     &\nodata  &\nodata     & 1800           \\
d062.wcc9\_3   & \nodata &  2003-10-29.26  & VLT       & AGN    & 2.42 &\nodata&\nodata & \nodata     &\nodata  & 20.3       & 1800           \\
d083.wdd9\_12  & 2003jn  &  2003-10-29.29  & VLT       & Ia     &\nodata & 0.33 & -1     & 91T  & good    & 20.8       & 1800           \\
d084.wdd9\_11  & 2003jm  &  2003-10-30.19  & VLT       & Ia     & 0.522 & 0.52  & +8     & 72E  & good    & 22.9       & 1800           \\
d085.waa5\_16  & 2003jv  &  2003-10-28.37  & GMOS      & Ia     & 0.405 & 0.41  & +3     & 94ae & poor    & 22.2       & 3x1200         \\
d086.waa5\_3   & 2003ju  &  2003-10-27.06  & GMOS      & Ia     &\nodata& 0.20  & -7    & 89B  & poor    & 21.6       & 3x600          \\
d086.waa5\_3   & 2003ju  &  2003-11-27.10  & Baade     & Ia     &\nodata& 0.20  & +13    & 89B  & good    & 21.6       & 3x1800         \\
d087.wbb5\_4   & 2003jr  &  2003-11-01.18  & GMOS      & Ia     & 0.340 & 0.34  & +6     & 95E  & good    & 21.9       & 3x600          \\
d089.wdd6\_8   & 2003jl  &  2003-10-31.34  & VLT       & Ia     & 0.429 & 0.43  & +6     & 95E  & good    & 22.4       & 1800           \\
d091.wcc1\_2   & \nodata &  2003-10-29.22  & VLT       & Unk    &\nodata&\nodata&\nodata & \nodata     &\nodata  &\nodata     & 2x1800         \\
d093.wdd5\_3 {\tablenotemark{o}}  & 2003js  &  2003-10-29.96  & VLT & Ia     & 0.363 & 0.36      & -6    & 90N  & good    & 22.0 & 923+600      \\
e142.wdd5\_3 {\tablenotemark{o}}  & 2003js  &  2003-11-23.21  & VLT & Ia     & 0.363  & 0.36     & +12   & 95D  & good    & 22.0 & 3x1200       \\
d097.wdd5\_10  & 2003jt  &  2003-10-29.32  & VLT       & Ia     &\nodata & 0.45 & -5     & 90O  & good    & 22.0       & 1800           \\
d099.wcc2\_16  & 2003ji  &  2003-11-01.23  & GMOS      & Ia     &\nodata& 0.21  & +17    & 95bd & good    & 20.9       & 3x600          \\
d100.waa7\_16  & 2003jq  &  2003-10-24.21  & FLWO      & Ia     &\nodata& 0.16  & +26    & 95al & good    & 19.8       & 2x1800         \\
d115.wbb6\_11  & \nodata &  2003-10-28.43  & GMOS      & Unk    &\nodata&\nodata&\nodata & \nodata     &\nodata  & 20.2       & 1200           \\
d117.wdd8\_16  & 2003jw  &  2003-10-30.32  & VLT       & Ia     & 0.296 & 0.29  & -1     & 95E  & good    & 22.6       & 1800           \\
d123.wcc9\_16  & \nodata &  2003-10-30.27  & VLT       & Gal    & 0.500 &\nodata&\nodata & \nodata     &\nodata  &\nodata     & \nodata  \\
d124.wcc9\_15  & \nodata &  2003-10-31.26  & VLT       & AGN    & 0.609 &\nodata&\nodata & \nodata     &\nodata  & 20.5       & 1800           \\
d149.wcc4\_11  & 2003jy  &  2003-10-31.10  & VLT       & Ia     & 0.339 & 0.34  & -5     & 90O  & good    & 22.7       & 1800           \\
d150.wcc1\_12  & \nodata &  2003-10-31.31  & VLT       & Gal    & 0.190 &\nodata&\nodata & \nodata     &\nodata  &\nodata     &\nodata         \\
d156.wcc2\_4   & 2003jx  &  2003-10-31.15  & VLT       & Unk    &\nodata&\nodata&\nodata & \nodata     &\nodata  &\nodata     & 2x1800         \\
e018.wbb7\_2   & \nodata &  2003-11-19.05  & Clay      & AGN    & 0.181 &\nodata&\nodata &  \nodata    &\nodata  & 18.6       & 600            \\
e020.waa6\_9   & 2003kk  &  2003-11-19.11  & Clay      & Ia     & 0.164 & 0.16  & -5     & 90O  & good    & 20.3       & 3x300          \\
e022.wbb7\_12  & 2003kj  &  2003-11-22.03  & VLT       & II     & 0.074 &\nodata&\nodata &\nodata      &\nodata  & 22.3       & 1800+900       \\
e025.wdd3\_15  & \nodata &  2003-11-19.21  & Clay      & Gal    & 0.180 &\nodata&\nodata & \nodata     &\nodata  &\nodata     & 3x1200         \\
e027.wcc7\_16  & \nodata &  2003-11-21.16  & VLT       & Unk    &\nodata&\nodata&\nodata & \nodata     &\nodata  & 22.5       & 3x1200         \\
e029.wbb3\_15 {\tablenotemark{p}} & 2003kl  &  2003-11-19.14  & Clay & Ia     & 0.335  & 0.33    & -5    & 90O  & good    & 21.0  & 3x600       \\
e121.wbb3\_15 {\tablenotemark{p}} & 2003kl  &  2003-11-22.11  & Clay & Ia     & 0.335  & 0.33    & -1    & 81B  & good    & 21.0  & 3x1200      \\
e103.wbb9\_2   & \nodata &  2003-11-21.05  & VLT       & Unk    & 0.871 &\nodata&\nodata & \nodata     &\nodata  &\nodata     & 1800           \\
e106.wbb6\_11  & \nodata &  2003-11-20.14  & Clay      & Unk    & 0.321 &\nodata&\nodata & \nodata     &\nodata  & 19.4       & 3x1200         \\
e108.wdd8\_4   & 2003km  &  2003-11-20.21  & Clay      & Ia     &\nodata& 0.47  & -8     & 90N  & good    & 21.8       & 3x1200       \\
e108.wdd8\_4   & 2003km  &  2003-11-21.21  & VLT       & Ia     &\nodata& 0.47 & -8    & 90N  & good    & 21.8       & 2x1800          \\
e118.waa5\_11  & \nodata &  2003-11-22.03  & VLT       & AGN    & 0.556 &\nodata&\nodata & \nodata     &\nodata  &\nodata     & 2x1200         \\
e119.wbb1\_7   & \nodata &  2003-11-23.19  & VLT       & Gal    & 0.560 &\nodata&\nodata & \nodata     &\nodata  &\nodata     & 1800           \\
e120.waa5\_9   & \nodata &  2003-11-22.05  & Clay      & Gal    & 0.298 &\nodata&\nodata & \nodata     &\nodata  &\nodata     & 1200           \\
e132.wcc1\_7   & 2003kn  &  2003-11-22.08  & VLT       & Ia     & 0.244 & 0.24  & -6     & 90N  & good    & 21.3       & 2x1800         \\
e133.wcc1\_7   & \nodata &  2003-11-22.08  & VLT       & Gal    & 0.244 &\nodata&\nodata & \nodata     &\nodata  &\nodata     & 2x1800         \\
e136.wcc1\_12  & 2003ko  &  2003-11-22.13  & VLT       & Ia     & 0.360 & 0.36  & -11    & 94D  & good    & 21.7       & 2x1800         \\
e138.wdd4\_1   & 2003kt  &  2003-11-23.10  & VLT       & Ia     &\nodata & 0.61 & +5     & 95D  & good    & 22.8       & 3x1200         \\
e140.wdd5\_15  & 2003kq  &  2003-11-22.29  & VLT       & Ia     & 0.606 & 0.62  & -8     & 90N  & good    & 22.6       & 3x1200         \\
e141.wdd7\_2   & \nodata &  2003-11-22.16  & Clay      & II     & 0.099 &\nodata&\nodata & \nodata     &\nodata  &\nodata     & 2x1200         \\
e143.wdd7\_3   & \nodata &  2003-11-23.15  & VLT       & Unk    &\nodata &\nodata&\nodata & \nodata     &\nodata  &\nodata     & 2x1200         \\
e147.wdd5\_9   & 2003kp  &  2003-11-22.18  & VLT       & Ia     &\nodata & 0.64 & -7     & 89B  & good    & 22.1       & 2x1800         \\
e148.wdd5\_10  & 2003kr  &  2003-11-22.23  & VLT       & Ia     & 0.427 & 0.42  & -7     & 90N  & good    & 22.0       & 3x1200         \\
e149.wdd5\_10  & 2003ks  &  2003-11-23.28  & VLT       & Ia?    &\nodata & 0.51 & +12    & 95bd & good    & 22.2       & 3x1200         \\
e309.waa9\_14  & \nodata &  2003-11-23.32  & GMOS      & M-star &\nodata&\nodata&\nodata & \nodata     &\nodata  &\nodata     & 3x1200         \\
e315.wbb9\_3   & 2003ku  &  2003-11-24.31  & GMOS      & Ia?    &\nodata& 0.79  & +8     & 72E  & poor    & 22.9       & 3x1200    \\ 
e418.wcc2\_8   & \nodata &  2003-11-27  & Baade     & N.S.    &\nodata&\nodata&\nodata & \nodata     &\nodata  &\nodata     & \nodata   \\
e501.waa1\_1   & \nodata &  2003-11-28  & Baade     & N.S.    &\nodata&\nodata&\nodata & \nodata     &\nodata  &\nodata     & \nodata        \\
e504.waa3\_4   & \nodata &  2003-11-29.05  & Baade     & AGN    & 0.674 &\nodata&\nodata & \nodata     &\nodata  & 23.2       & 3x1800         \\
e510.waa1\_13  &\nodata  &  2003-11-29.08  & GMOS      & Unk    &\nodata&\nodata&\nodata & \nodata     &\nodata  & 23.0       & 1800           \\
e528.wcc5\_3   & \nodata &  2003-11-28  & Baade     & N.S.   &\nodata&\nodata&\nodata & \nodata     &\nodata  & 23.4       & \nodata        \\
e529.wcc5\_3   &\nodata  &  2003-11-29.10  & Clay      & Unk    &\nodata&\nodata&\nodata &\nodata & \nodata & 23.8       & 3x1800         \\
e531.wcc1\_4   & 2003kv  &  2003-11-29.14  & Baade     & Ia?    &\nodata& 0.78  & -3     & 95E  & poor    & 23.4       & 3x1800         \\
f001.wbb7\_1   & 2003lg  &  2003-12-19.17  & MMT       & Unk    &\nodata&\nodata&\nodata & \nodata     &\nodata  & 22.5       & 3x1800         \\
f011.wcc7\_12  & 2003lh  &  2003-12-21.31  & KI/LRIS   & Ia     &\nodata& 0.54  & +4     & 90N  & good    & 22.7       & 1500           \\
f017.wbb9\_10  & \nodata &  2003-12-20.33  & GMOS      & AGN    & 0.725 &\nodata&\nodata & \nodata     &\nodata  & 22.7       & 3x1200         \\
f041.wbb6\_8   & 2003le  &  2003-12-20.23  & KI/LRIS   & Ia     &\nodata& 0.56  & -4     & 94D  & good    & 22.7       & 2x1200         \\
f044.wbb8\_8   &\nodata  &  2003-12-21.10  & MMT       & Gal    & 0.409 &\nodata&\nodata & \nodata     &\nodata  &\nodata     & 3x1800         \\
f076.wbb9\_01  & 2003lf  &  2003-12-21.29  & KI/LRIS   & Ia     &\nodata& 0.41  & +3     & 94ae & good    & 22.1       & 900            \\
f076.wbb9\_1   & 2003lf  &  2003-12-21.17  & MMT       & Ia     &\nodata& 0.41  & +4     & 95D  & good    & 22.1       & 3x900          \\
f095.wcc2\_8   &\nodata  &  2003-12-20.31  & KI/LRIS   & Gal    & 0.313 &\nodata&\nodata & \nodata     &\nodata  & 21.6       & 3x1200         \\
f096.waa3\_3   & 2003lm  &  2003-12-21.28  & KI/LRIS   & Ia     & 0.408 & 0.41  & -1     & 92A  & good    & 22.5       & 1500           \\
f116.wbb1\_7   &\nodata  &  2003-12-20     & KI/LRIS   & N.S.   &\nodata&\nodata&\nodata &  \nodata    &\nodata  & 21.6       &  \nodata              \\
f123.wcc1\_7   &\nodata  &  2003-12-21.34  & KI/LRIS   & Gal    & 0.526 &\nodata&\nodata & \nodata     &\nodata  & 21.6       & 2x1200         \\
f213.wbb4\_12  & \nodata &  2003-12-19.34  & GMOS      & Unk    &\nodata&\nodata&\nodata & \nodata     &\nodata  &\nodata     & 2x1200         \\
f216.wdd4\_15  & 2003ll  &  2003-12-21.44  & KI/LRIS   & Ia     & 0.596 & 0.60  & +5     & 95E  & good    & 21.6       & 1800           \\
f221.wcc4\_14  & 2003lk  &  2003-12-21.36  & KI/LRIS   & Ia?    & 0.442 & 0.45  & +2     & 94ae & poor    & 22.8       & 1500           \\
f231.waa1\_13  & 2003ln  &  2003-12-21.25  & KI/LRIS   & Ia     &\nodata& 0.63  & +0     & 81B  & good    & 22.9       & 1500           \\
f235.wbb5\_13  & 2003lj  &  2003-12-20.27  & KI/LRIS   & Ia     & 0.417 & 0.42  & +5     & 95D & good    & 22.1       & 2x1200         \\
f244.wdd3\_8   & 2003li  &  2003-12-20.42  & KI/LRIS   & Ia     & 0.544 & 0.54  & -1     & 95E  & good    & 22.8       & 2x1800         \\
f301.wdd6\_1   &\nodata  &  2003-12-21.42  & KI/LRIS   & Ia?    &\nodata&0.52&-3      & 86G  &poor  & 21.6       & 1500           \\
f304.wdd6\_2   &\nodata  &  2003-12-21.39  & KI/LRIS   & Unk    &\nodata&\nodata&\nodata & \nodata     &\nodata  & 21.6       & 1800           \\
f308.wdd6\_10  &\nodata  &  2003-12-20.37  & KI/LRIS   & Ia?    &\nodata& 0.39  & -7     & 94D  & poor    & 21.6       & 3x1800         \\
f441.wbb6\_7   &\nodata  &  2003-12-23.25  & GMOS      & Unk    &\nodata&\nodata&\nodata & \nodata     &\nodata  &\nodata     & 3x1200         \\
\enddata                        

\tablenotetext{a}{ESSENCE internal identification.  The first letter
  indicates the month in the observing season.  This is followed by a
  sequential number as targets are discovered.  The remaining letters
  and numbers show the specific ESSENCE field where the object was
  located \citep{smith05}.}
\tablenotetext{b}{Note that not all objects judged to be SNe have
  official International Astronomical Union names.}
\tablenotetext{c}{The UT date at the midpoint of the observation(s).}
\tablenotetext{d}{Our best guess as to classification of the object.
  Ia? indicates a lack of certainty in the identification as a SN~Ia.
  II? indicates a lack of certainty in the identification as a SN~II.
  Objects marked Unk are unknown.  N.S. indicates that the
  telescope was pointed to the object, but no exposure was taken or
  the exposure contained no signal.}

\tablenotetext{e}{Redshift measured from narrow emission or absorption
  lines from the host galaxy.}
\tablenotetext{f}{Redshift measured from the SN spectrum by SNID.}
\tablenotetext{g}{Age of the SN relative to maximum brightness based
  upon comparisons with SNID templates.}
\tablenotetext{h}{Template spectrum in SNID that provides the best
  match to the observed spectrum.}
\tablenotetext{i}{Qualitative judgment about the SNID fit.}
\tablenotetext{j}{Magnitude at discovery, not at time of spectroscopy.}

\tablenotetext{k}{As a result of problems with the LRIS spectrograph,
  b027 was observed for 1800 s on the blue side only, followed by a
  gap of a little over half an hour, then observed for 1800 s with
  both the blue and red sides.  The time of observation listed is
  halfway between the midpoints of the two observations.}
\tablenotetext{l}{The SNID analysis was performed on the weighted
  combination of the GMOS and Clay spectra.}
\tablenotetext{m}{Objects d009 and d106 are the same, inadvertently
  assigned two different internal identifications.}
\tablenotetext{n}{SN~1987M is of Type Ic.}
\tablenotetext{o}{Objects d093 and e142 are the same, inadvertently
  assigned two different internal identifications.}
\tablenotetext{p}{Objects e029 and e121 are the same, inadvertently
  assigned two different internal identifications.}
\end{deluxetable}                                
\clearpage

\end{document}